\documentclass{aa}
\usepackage{graphicx}
\begin{document}
   \title{Origin of the 6.85 $\mu$m Band near 
Young Stellar Objects: The Ammonium Ion ($\rm NH_4^+$) Revisited
\thanks{Based on observations with ISO, an ESA project with instruments
funded by ESA Member States (especially the PI countries: France, Germany,
the Netherlands and the United Kingdom) and with the participation of ISAS
and NASA}}

\titlerunning{$\rm NH_4^+$ and the 6.85 $\mu$m band near Young Stellar Objects}

\author{W. A. Schutte\inst{1} \and R. K. Khanna  \inst{2}}

\authorrunning{W. A. Schutte et al.}

\offprints{W. A. Schutte}
\institute{Raymond and Beverly Sackler Laboratory for Astrophysics, Leiden
Observatory, P.O. Box 9513, 2300 RA Leiden, The Netherlands
\and Department of Chemistry and Biochemistry, University of Maryland,
College Park, MD 20742, USA}

\date{Received - , 2000; Accepted - , 2000}

\abstract{We have investigated whether the $\nu_4$ feature of $\rm
NH_4^+$ is a viable candidate for the 6.85 $\mu$m absorption band seen
towards embedded young stellar objects. To produce $\rm NH_4^+$
astrophysical ice analogs consisting of $\rm H_2O$, $\rm CO_2$, $\rm
NH_3$ and $\rm O_2$ were UV photolysed. The IR spectra reveal peaks
that are identified with the NH$_4^+$, NO$_2^-$, NO$_3^-$ and
HCO$_3^-$ ions. It is shown that the $\rm NH_4^+$
matches two absorption features that are observed towards embedded
young stellar objects, i.e., the strong 6.85 $\mu$m feature and the
3.26 $\mu$m feature.  The characteristic redshift with temperature of
the interstellar 6.85 $\mu$m feature is well reproduced. The abundance
of $\rm NH_4^+$ in interstellar ices would be typically 10 \% relative
to $\rm H_2O$.  The experiments show that the counterions produce
little distinct spectral signature but rather a pseudo-continuum if a
variety of them is present in a $\rm H_2O$ dominated environment. The
anions could therefore go undetected in IR spectra of interstellar
ice. In the ISM, where additional mechanisms such as surface chemistry
and additional elements such as sulfur are available many acids and an
even wider variety of anions could be produced.  These components may
be detectable once the ices sublime, e.g, in hot cores.

}
   \maketitle

%
%________________________________________________________________

\section{Introduction}

The nature of the 6.85 $\mu$m absorption feature towards embedded
Young Stellar Objects (YSO's) has remained an enigma since its
discovery 25 years ago (Russell et al. 1977). While the first high
resolution observations of this band were recently obtained by the
Short Wavelength Spectrometer (SWS) on board the Infrared Space
Observatory (ISO; Schutte et al. 1996, Dartois et al. 1999a; Keane et
al. 2001), this did not yet help to clarify its origin. A number of
candidates, such as carbonates, and the CH deformation modes in
organic molecules like methanol, could be excluded based on the
feature's spectral properties and the absence of strong additional
bands. One possibility, however, the $\nu_4$ mode of the ammonium ion
($\rm NH_4^+$), first proposed by Grim et al. (1989b), could not be
excluded. $\rm NH_4^+$ is produced in astrophysical ice analogs by
acid-base reactions.  This is achieved either by deposition and
warm-up of $\rm NH_3$ together with acids such as HNCO or HCOOH
(Novozamsky et al. 2001, Schutte et al. 1999), or by photolysis of ice
mixtures containing $\rm NH_3$ (e.g., $\rm NH_3$/$\rm O_2$; Grim et
al. 1989a) where the acids are produced by the
photochemistry. Recently, the good spectral correspondence between the
$\nu_4$ feature of $\rm NH_4^+$ and the interstellar 6.85 $\mu$m
absorption as seen by ISO/SWS was again demonstrated (Demyk et
al. 1998). Nevertheless, this assignment faces a fundamental problem,
since it requires a large abundance of $\rm NH_4^+$ of $\sim$ 10 \% in
the ices near YSO's. An equal amount of negative charge would need to
be present. Although the ions $\rm OCN^-$ (also referred to as NCO- in
the chemistry literature) and (probably) $\rm HCOO^-$ have been
identified, their abundance falls far short of what is needed for the
balance (Schutte et al. 1996a; Gibb et al. 2000; Keane et al. 2001;
2002). No sign of other negative solid state species has shown up in
the ISO data.

This paper revisits the possibility that $\rm NH_4^+$ is responsible
for the interstellar 6.85 $\mu$m band. We investigate in the
laboratory whether a {\it variety} of anions can supply the required
countercharge, and that such a combination of species is either
lacking in outstanding IR features, or fall below current detection
limits. Taking the cosmic abundances of the elements into
consideration, the counterions would have to consist of H, O, C, N and
S in order to make a significant contribution. If we furthermore
constrain the acid precursors to species with at most 1 carbon atom,
possibilities would still include a variety of thermodynamically
stable species, i.e., OH$^-$, CN$^-$, OCN$^-$, HCOO$^-$, $\rm HCO_3^-$, $\rm
CO_3^{2-}$, NO$_2^-$, NO$_3^-$, $\rm SO_4^{2-}$, HSO$_3^-$, and $\rm
SO_3^{2-}$. In principle an astrophysical ice analog containing $\rm
NH_4^+$ and a mixture of such negative ions could be prepared by
depositing $\rm NH_3$ together with a variety of
acids. However, experimental limitations compel us to produce the ions
{\it in situ} with UV photolysis.  To this end, we processed mixtures of
$\rm H_2O$/$\rm CO_2$/ $\rm NH_3$/$\rm O_2$ to produce acids such as
$\rm H_2CO_3$ (carbonic acid), $\rm HNO_3$ (nitric acid) and $\rm
HNO_2$ (nitrous acid). These acids react with $\rm NH_3$ forming $\rm
NH_4^+$ and a mixture of counterions, as desired. Earlier work on 
similar mixtures
indeed showed the formation of such species (Moore \& Khanna 1991;
Gerakines et al. 2000; Grim et al. 1989b).  Of course, this
experimental method of producing the ions may be different from what
happens in the ISM. For example, the high abundance of $\rm O_2$ in
our samples, necessary to stimulate the photochemical production of
oxygen rich acids, may considerably exceed the abundance in
interstellar ices (Vandenbussche et al. 1999). In space
acids are possibly produced by other mechanisms such as surface
chemistry (e.g., Keane \& Tielens 2002, in preparation).

To verify whether $\rm NH_4^+$ is a plausible candidate for the identification
of the 6.85 $\mu$m band a number of
crucial issues need to be adressed. First of all,
detailed spectroscopy of the $\nu_4$ mode is needed to 
investigate its behaviour as a function of ice composition and temperature.
Next, the issue whether other infrared features of $\rm NH_4^+$ 
could be present in interstellar ice spectra must be adressed. Subsequently,
the counterions 
that are produced in the experiments should be identified to assess their
astrophysical relevance. Finally, the spectral signature of the counterions
under various conditions is studied, where the prime question is whether
conditions exist at which their infrared signatures become inconspicuous. 

The paper is organized as follows. In Sect. 2 we review the
experimental techniques. Sect. 3 summarizes the results. $\rm NH_4^+$
is created by UV photolysis of astrophysical ice analogs. We
describe the spectral properties of $\rm NH_4^+$ under such
conditions.  Furthermore we
establish which counterions are formed in the experiments and study
their spectroscopic properties as well. In sect. 4 the obtained $\rm NH_4^+$
spectra are compared with observations of YSO's. Besides matching
the 6.85 $\mu$m band, it is furthermore
shown that the interstellar 3.26 $\mu$m feature closely corresponds to one
of the $\rm NH_4^+$ features. Also, using the
experimental results, it is investigated whether the absence of
features due to counterions in the observations can be reconciled with
an assignment of the 6.85 $\mu$m band to $\rm NH_4^+$. In Sect. 5, the
astrophysical implications of the $\rm NH_4^+$ identification are
discussed.  Sect. 6, finally, summarizes the conclusions of this
paper.

\section{Experimental}

\label{experimental}

Detailed descriptions of the general procedure for creation and
photolysis of ice samples and the measurement of their infrared
spectra have been published earlier (Gerakines et al. 1995; 1996).
In summary, the set-up consists of a high vacuum chamber ($\rm
10^{-7}$ mbar), with an IR transparent CsI substrate mounted on a cold
finger which is cooled to $\sim$ 12 K. Samples were slowly deposited
($\sim$ 3 $\times$ 10$^{15}$ molec. $\rm cm^{-2}$ $\rm s^{-1}$/4
$\mu$m hr$^{-1}$) through a narrow tube controlled by a regulation
valve. Photolysis by vacuum UV was subsequently performed by a
hydrogen flow discharge lamp ($\sim$ 5 $\times$ 10$^{14}$ photons $\rm
cm^{-2}$ s$^{-1}$; E$_{photon}$ $\geq$ 6 eV). For thorough photolysis,
the thickness of the sample should be $\sim$$<$ 0.2 $\mu$m.  To
overcome this limit, in one case deposition and photolysis were
performed simultaneously. In this way a photolysed sample can be
produced of several microns thickness, which greatly enhances the
amount of photoproducts and thus the S/N ratio of the IR
spectrum. Afterwards the sample was warmed in steps. The evolution of
the sample throughout the photolysis and warm-up sequence was
monitored by infrared transmission spectroscopy. In addition to the
photochemical experiments, we did band strength measurements for $\rm
NH_4^+$ by deposition and warm-up of ices containing $\rm NH_3$ and
HCOOH.

The reagents used in these experiments were H$_2$O (purified by three 
freeze-thaw cycles), $\rm CO_2$ (Praxair, 99.996$\%$ purity),
$\rm NH_3$ (Praxair, 99.99 $\%$ purity), $\rm O_2$ (Praxair,
99.999 $\%$ purity) and HCOOH (Baker, 98 \%). When ice samples containing both 
$\rm NH_3$ and $\rm CO_2$ or $\rm NH_3$ and HCOOH
 were prepared, the $\rm CO_2$ or HCOOH was deposited through a separate 
tube, to prevent reactions with $\rm NH_3$ prior to deposition.
This method is described in Gerakines et al. (1995).

%A detailed discussion of the laboratory set-up for the proton
%irradiation of ice at NASA/Goddard Space Flight Center has been
%published (Hudson \& Moore 1999, and references therein). In summary,
%a gas mixture of $\rm H_2O$ and $\rm CO_2$ was prepared inside a
%vacuum line separate from the vacuum manifold containing $\rm NH_3$
%gas. Using two different capillary deposition tubes, the $\rm H_2O$ +
%$\rm CO_2$ mixture and $\rm NH_3$ gas were simultaneously condensed
%onto a cold (18 K) aluminum mirror suspended inside a stainless-steel
%high-vacuum chamber (P $\sim$ 10$^{-7}$ mbar). IR spectra were
%recorded (60-scan accumulations at a resolution of 4 $\rm cm^{-1}$) by
%diverting the beam of an FTIR spectrometer toward the ice-covered
%mirror, where it passed through the ice before and after reflection at
%the ice-mirror interface. Facing the mirror toward a beam of 0.8 MeV
%protons generated by a Van de Graaff accelerator irradiates the
%ice. The reagents used were: $\rm H_2O$ (triply distilled, with a
%resistance greater than 10$^7$ ohm cm), $\rm CO_2$ (Matheson, 99.995
%\%) and $\rm NH_3$ (99.99 \%, anhydrous, Matheson).

\begin{table*}
\caption[]{Log of experiments}
\begin{flushleft}
\label{log}
\begin{tabular}{lllllcrcc}
\hline
& \multicolumn{4}{c}{Mixture} &\multicolumn{1}{c}{N} & 
\multicolumn{1}{c}{d$^a$} & \multicolumn{1}{c}{dose} &
\multicolumn{1}{c}{dose} \\
&H$_2$O & CO$_2$ & NH$_3$ & O$_2$ & molec. cm$^{-2}$ & $\mu$m & 
photons cm$^{-2}$ & eV molec.$^{-1}$$^d$ \\
\hline
1 & 10 & 10 & 0 & 10 & 7.2(17) & 0.37 & 9(17) & $<$11$^c$ \\ 
2 & 10 & 5 & 7 & 10 & 4.0(17) & 0.17 & 9(17) & $<$20$^c$ \\
3$^b$ & 10 & 2.2 & 1.2 & 1 & 2.5(19) & 9.5 & 2.4(19) & 8.6 \\
4 & 10 & 2 & 1.1 & 1 & 8.5(17) & 0.32 & 6(17) & $<$6.4$^c$ \\
5 & 10 & 1.2 & 0.9 & 0.9 & 4.8(17) & 0.17 & 9(16) & $<$1.7$^c$ \\
6 &  0 & 1.2 & 1 & 1 & 4.4(17) & 0.23 & 9(16) & $<$1.8$^c$ \\
7 & 0 & 0 & 1 & 1 & 4.5(17) & 0.18 & 9(17) & $<$18$^c$ \\
%8 & 10 & 2 & 1.5 & 1 & 1.1(19) & 3.6 & - & 10 \\
\hline
\end{tabular}
\end{flushleft}
\begin{list}{}{}
%\item[$^{\rm a}$] Expts. 1 - 7: Samples photolysed by UV; Expt. 8:
%Sample irradiated with 0.8 MeV protons.
\item[$^{\rm a}$] Thickness obtained from the column density using a nominal 
density of 1 g cm$^{-3}$.
\item[$^{\rm b}$] Obtained by simultaneous deposition and photolysis.
\item[$^{\rm c}$] Sample not optically thick, therefore dose is an upper limit.
\item[$^{\rm d}$] Using an average photon energy of 9 eV (Jenniskens et al.
1993). 
%For expt. 8, the dose was estimated assuming a stopping power
%of 380 MeV $\rm cm^{-1}$ (Hudson and Moore 1998). 
\end{list}
\end{table*}

Table~\ref{log} provides a log of our experiments.  Column densities
and abundances of $\rm H_2O$, $\rm CO_2$ and $\rm NH_3$ were directly
obtained from the IR spectrum using band strengths from the literature
(Gerakines et al. 1995; Kerkhof et al. 1999).

\section{Results}

%\subsection{UV photolysis}

%\subsubsection{$\rm H_2O$/$\rm CO_2$/$\rm O_2$ mixtures}

%   \begin{figure}
%   \centering
%   \includegraphics[width=9.4cm]{amca_fig1.eps}
%\caption{Spectrum of the photolysed ice sample 
%$\rm H_2O$/$\rm CO_2$/$\rm O_2$ = 1/1/1 (expt. 1); a. at 12K; b. 
%after warm-up 
%to 220K. The spectra have been offset for clarity.
%              }
%         \label{comp3part}
%   \end{figure}

%In preparation to the full four component mixture,
%we first study the photochemistry of $\rm H_2O$/$\rm CO_2$/$\rm O_2$ ices.
%The reason for this is that in the absence of ammonia no acid-base
%reactions will occur, allowing a look at the original acid precursors.
%This will give better constraints for the identification of the
%negative ions that are produced once $\rm NH_3$ is added.
%
%Fig.~\ref{comp3part} presents the 1900 - 1000 cm$^{-1}$ spectrum of
%$\rm H_2O$/$\rm CO_2$/$\rm O_2$ = 1.0/1.0/1.0 after photoprocessing
%(Expt. 1, Table~\ref{log}), showing clear new bands at 1727, 1480 and
%1275 $\rm cm^{-1}$ next to the 1650 $\rm cm^{-1}$ $\rm H_2O$ bending
%mode.  After warm-up to 220 K and sublimation of the original ices,
%these features persist while shifting to 1713, 1505 and 1304 $\rm
%cm^{-1}$.  Other strong features are found at 3040,
%2840, and 2627 $\rm cm^{-1}$. These bands demonstrate the formation of
%carbonic acid ($\rm H_2CO_3$; Gerakines et al. 2000).

\subsection{Photolysis of $\rm H_2O$/$\rm CO_2$/$\rm NH_3$/$\rm O_2$ mixtures}

\begin{figure}
   \centering
   \includegraphics[width=9.4cm]{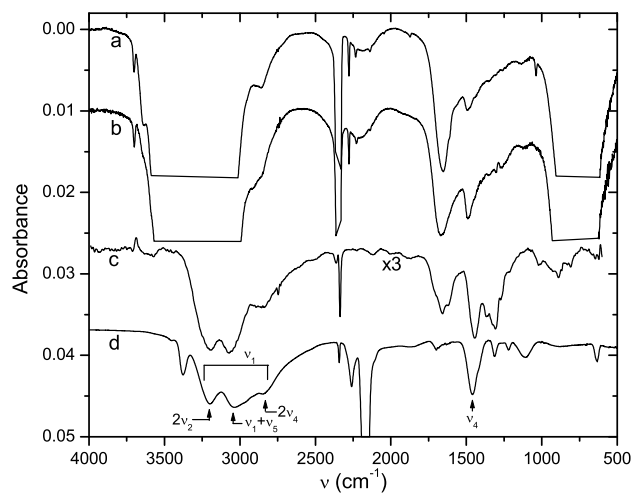}
      \caption{Spectra of the photolysed sample 
$\rm H_2O$/$\rm CO_2$/$\rm NH_3$/$\rm O_2$=
10/2/1.1/1 (expt. 4): a. Before warm-up (12 K); b. at 120 K;
c. at 220 K. Curve c was scaled up by a factor 3. 
Curve d shows the spectrum of HNCO/$\rm NH_3$ =
1/1.2 at 120 K (after the transformation
of most of the original species into $\rm OCN^-$ and $\rm NH_4^+$;
arrows identify features of $\rm NH_4^+$). The spectra have
been offset for clarity.}
         \label{comp4}
   \end{figure}

We photolysed a number of mixtures of 
$\rm H_2O$/$\rm CO_2$/$\rm NH_3$/$\rm O_2$ to produce $\rm NH_4^+$
and a variety of negative ions (Table~\ref{log}).
Fig.~\ref{comp4} shows the evolution of spectrum of photolysed
$\rm H_2O$/$\rm CO_2$/$\rm NH_3$/$\rm O_2$ = 10/2/1.1/1
(Expt. 4; Table~\ref{log}) during warm-up from 12 K to 220 K.
First of all we want to verify whether $\rm NH_4^+$ is formed.
Therefore, Fig.~\ref{comp4}
also shows the spectrum of the binary mixture HNCO/$\rm NH_3$ =
1/1.2, after warm-up to 120 K (no photolysis).  In this sample $\rm
NH_4^+$ is readily formed by proton transfer between the isocyanic acid
and the ammonia (Novozamsky et al. 2001). The $\rm NH_4^+$ features fall
at 3000 $\rm cm^{-1}$ (broad; $\nu_1$), 3200 $\rm cm^{-1}$ (2$\nu_2$), 3060 
$\rm cm^{-1}$
($\nu_1$+$\nu_5$), near 2860 $\rm cm^{-1}$ ($2\nu_4$) and near 1450
$\rm cm^{-1}$ ($\nu_4$), where $\nu_5$ is a lattice mode (assignments from 
Nakamoto 1972). The first four
features blend together in a broad structure which extends from $\sim$
3500 - 2400 $\rm cm^{-1}$. Directly after the photolysis of the $\rm
H_2O$/$\rm CO_2$/$\rm NH_3$/$\rm O_2$ mixture, the $\nu_4$ feature is
already clearly present at $\sim$ 1500 $\rm cm^{-1}$. 
The H$_2$O ice absorption in the 3400- 2800 cm$^{-1}$ region obscures the other
absorptions due to NH$_4^+$. Upon warming to 220 K, causing the ice to
sublime, the other $\rm
NH_4^+$ features can all be seen. Clearly, the ammonium ion is produced
by the photolysis.

\begin{figure}
   \centering
   \includegraphics[angle=-90,width=8.8cm]{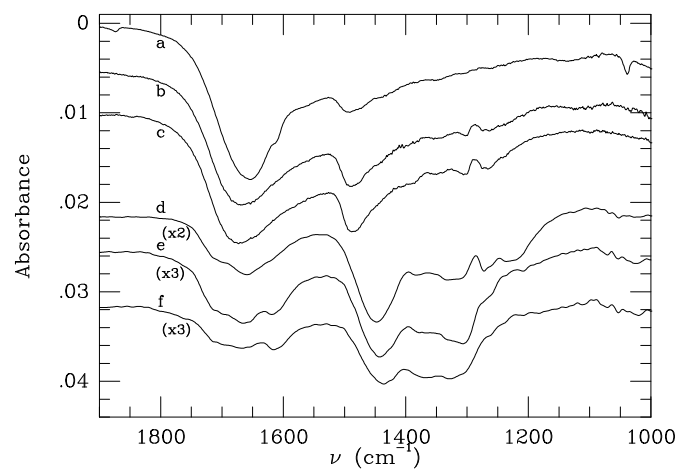}
      \caption{Thermal evolution of the photolysed ice mixture 
$\rm H_2O$/$\rm CO_2$/$\rm NH_3$/$\rm O_2$=
10/2/1.1/1 (expt. 4), blow-up of the 1900 - 1000 $\rm cm^{-1}$ region: 
a. before warm-up
(12 K); b. after warm-up to 120 K; c. to 150 K; d. to 180 K, e. to 220 K;
f. to 240 K. Apart from the multiplication factors indicated in the
figure, the vertical scale is identical for all
curves. However, the spectra have been offset for clarity.}
         \label{comp4part}
   \end{figure}

   \begin{figure}
   \centering
   \includegraphics[angle=-90,width=8.8cm]{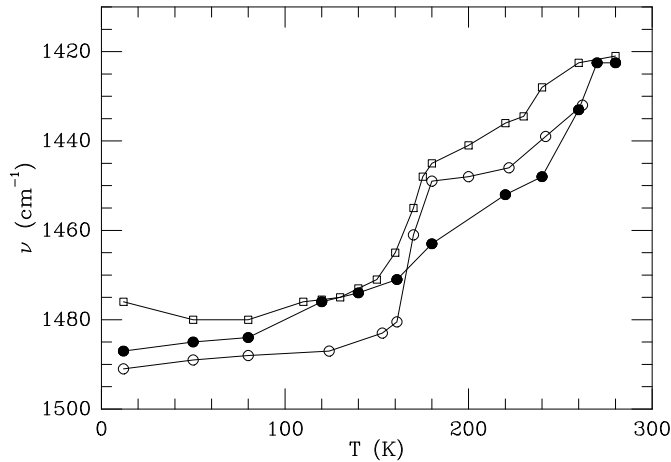}
\caption{Redshift of the $\nu_4$$\rm NH_4^+$ feature as a function
of temperature: $\rm H_2O$/$\rm CO_2$/$\rm NH_3$/$\rm O_2$ =
10/2/1.1/1 (expt. 4; open circles); 1/0.5/0.7/1 (expt. 2; open squares); 
0/1.2/1/1 (expt. 6; filled circles). 
              }
         \label{redshift}
   \end{figure}

As discussed in Sect. 1, we want to investigate the spectral properties
of the $\rm \nu_4$ $\rm NH_4^+$ feature.
To study the temperature dependence, Fig.~\ref{comp4part} shows 
the 1900 - 1000 $\rm
cm^{-1}$ spectra of photolysed $\rm H_2O$/$\rm CO_2$/$\rm NH_3$/$\rm
O_2$ = 10/2/1.1/1 (expt. 4). Spectra are shown directly after
photolysis at 12 K, and at a variety of temperatures up to 240 K. Some
important spectral properties can be gleaned from the figure. Upon
warm-up to 120 K the depth of the $\nu_4$ feature near 1460 $\rm
cm^{-1}$ increases by a factor $\sim$ 2. While this increase could
result from further acid-base reactions during the warm-up, the
absence of any feature attributable to acids such as $\rm H_2CO_3$ or 
$\rm HNO_3$
after the photolysis (Fig.~\ref{comp4part}, $\rm H_2CO_3$ has bands
at 1727, 1480 and 1275 $\rm cm^{-1}$; Gerakines et al. 2000;
$\rm HNO_3$ has a
strong feature at 1300 cm$^{-1}$; McGraw et al. 1965) argues
against this possibility.  Thus, the growth of the $\rm NH_4^+$
feature during warm-up is likely caused by an increase of its
intrinsic strength.  Furthermore, the $\nu_4$ feature shifts strongly
redward with temperature. The enhancement and the shift reflect the
strong interaction of the ion with its
environment. Fig.~\ref{redshift} plots the position of the $\nu_4$
band vs. temperature for a number of samples (expts. 2, 4 and 6 of
Table~\ref{log}).  It can be seen that, while there is a steady
redshift throughout the warm-up, the shift is particularly pronounced
during $\rm H_2O$ sublimation between 160 and 180 K.

   \begin{figure}
   \centering
   \includegraphics[width=9.4cm]{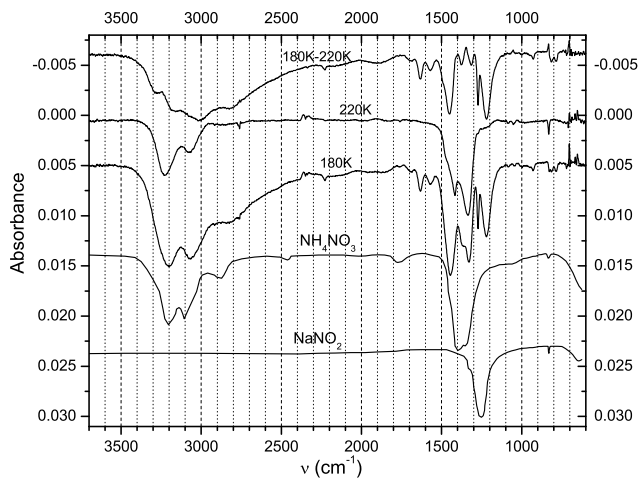}
   \caption{The thermal evolution of the residue of photolysed
$\rm NH_3$/$\rm O_2$ = 1/1 (Expt. 7). From central to top: 
the residue at 180 K; at 220 K; subtraction of the 220 K from the 180 K 
spectrum. The vertical scale is identical for all curves, however, the
spectra have been offset for clarity. For comparison, ammonium nitrate
($\rm NH_4NO_3$) and sodium nitrite (Na$\rm NO_2$) are shown at the bottom 
(Miller \& Wilkins 1952).
               }
              \label{comp2res}%
    \end{figure}

As discussed in Sect. 1, it is essential to identify the negative ions
produced in the experiments to assess their astrophysical
significance.  To this end, we study the IR spectrum of the residue
which remains after the ices have fully evaporated.  As a basis for
the interpretation of more complex samples, Fig.~\ref{comp2res} shows
the warm-up behaviour of the residue of photolysed $\rm NH_3$/$\rm
O_2$ = 1/1 (expt. 7; Table~\ref{log}). The evolution is best
understood by a ``hot to cold'' interpretation, i.e., starting with
the most refractory component at the final stages of the warm-up, and
then working backwards in the sublimation sequence to see which
spectral elements are sequentially lost.  After warm-up to 220 K, a
component remains which is stable up to $\sim$ 280 K. Its spectrum
resembles that of ammonium nitrate (Fig.~\ref{comp2res}; $\rm
NH_4NO_3$; throughout the paper we will use the standard chemical
notation for salts. It must be noted however that the constituents of
these compounds are ions, in this case, $\rm NH_4^+$ and $\rm
NO_3^-$). The $\rm NO_3^-$ ion was previously identified in photolysed
$\rm NH_3$/$\rm O_2$ ices (Grim et al.  1989b).  However, while there
is a general correspondence with the literature spectrum of ammonium
nitrate there are some differences in the relative intensities and
positions of the bands.  The literature spectrum was produced from
fine crystalline powder in nujol mull, while our samples consist of a
mixture of salts at low temperature.  Spectra of ions are generally
quite sensitive to factors like temperature, matrix and degree of
annealing, due to the strong interaction of the ions with the matrix
(c.f. OCN$^-$ in various salt matrices; Maki \& Decius 1959). This
effect possibly causes the differences. During warm-up from 180 to 220
K two strong features disappear at 1270 and 1220 $\rm cm^{-1}$,
together with a fraction of the 1450 $\rm cm^{-1}$, and 2400 - 3500
$\rm cm^{-1}$ complex due to $\rm NH_4^+$.  The band at 1220 $\rm
cm^{-1}$ may correspond to the main feature of $\rm NO_2^-$.  As
compared with the literature spectrum of NaNO$_2$ (in nujol mull;
Fig.~\ref{comp2res}), the band is shifted by 35 $\rm cm^{-1}$,
possibly due to matrix interactions. The assignment of this component
with ammonium nitrite ($\rm NH_4NO_2$) is supported by the
identification of the $\rm NO_2^-$ ion in the photolysed ice (Grim et
al.  1989b).  The 1270 $\rm cm^{-1}$ feature corresponds to an
unidentified photoproduct.

   \begin{figure*}
   \centering
   \includegraphics[width=16cm]{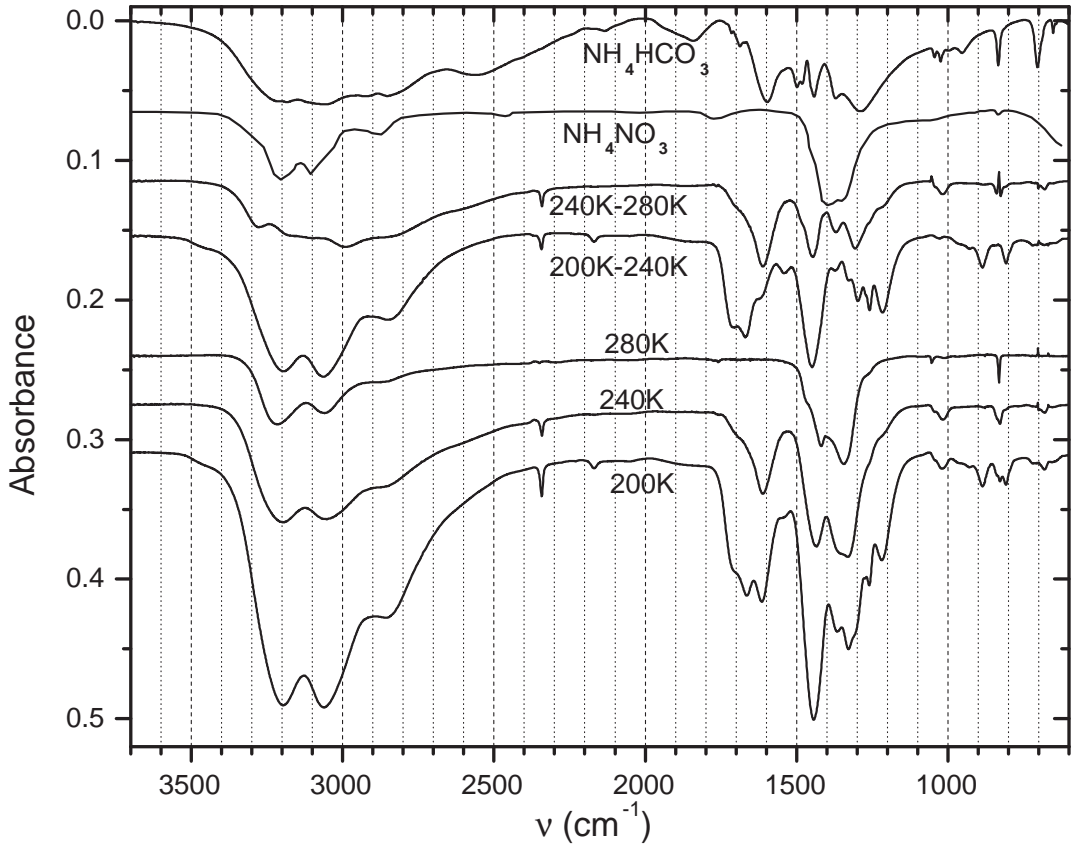}
   \vskip-30pt
   \caption{The thermal evolution of the residue of photolysed
$\rm H_2O$/$\rm CO_2$/$\rm NH_3$/$\rm O_2$ = 10/2.2/1.2/1 
(Expt. 3). From bottom to top:
The spectrum of the sample at 200 K; at 240 K; at 280 K; subtraction
of the 240 K from the 200 K spectrum; subtraction of the the 280 K
from the 240 K spectrum. The vertical scale is identical for all
curves, however, the spectra have been offset for clarity.  For
comparison, ammonium nitrate ($\rm NH_4NO_3$; Miller \& Wilkins
1952) and ammonium bicarbonate ($\rm NH_4HCO_3$; Khanna \& Moore
1999) are shown.
               }
              \label{comp4res}%
    \end{figure*}

Fig.~\ref{comp4res} shows the spectral evolution of the residue of
photolysed $\rm H_2O$/$\rm CO_2$/$\rm NH_3$/$\rm O_2$ = 10/2.2/1.2/1
(expt. 3). To probe the nature of the anions, we will again 
analyse this diagram ``hot to cold''. As with the $\rm NH_3$/$\rm O_2$ sample, 
the most refractory component,
remaining at 280 K after all other material has sublimed, resembles
ammonium nitrate (Fig.~\ref{comp4res}).  During warm-up from 240 to
280 K a component sublimes which is characterized by two strong bands
at 1600 and 1300 $\rm cm^{-1}$, together with four weaker features at
1375, 1010, 830, and 690 $\rm cm^{-1}$. The disappearance of these
bands is accompanied by a decrease of the $\rm NH_4^+$ features. All
these features have counterparts in the spectrum of
ammonium bicarbonate in a KBr pellet 
($\rm NH_4HCO_3$; Fig.~\ref{comp4res}). Still,
some differences are apparent in width and relative intensity, 
while furthermore some extra bands are present in the
literature spectrum of which only minor indications are seen
in the 240 to 280 K component. Again, we note that the spectra of ions
are generally quite sensitive to such factors as temperature, matrix
and degree of annealing, and it seems likely that the differences
could derive from such factors. 

To verify the presence of $\rm HCO_3^-$ in the 240 - 280 K fraction,
we photolysed the ice mixture $\rm H_2O$/$\rm CO_2$/$\rm O_2$ = 1/1/1
(Expt. 1, Table~\ref{log}) to see whether its precursor,
carbonic acid ($\rm H_2CO_3$) is formed when $\rm NH_3$ is omitted from the
mixture. The photolysis gave rise to
features at 3040, 2840, 2627, 1727, 1480, and 1275. These bands can all
be ascribed to carbonic acid (Moore \& Khanna 1991; Gerakines et al. 2000).

The residue fraction subliming between 200 - 240 K
is characterized by a rather complex spectrum.
%which, besides the $\rm
%NH_4^+$ bands, shows features at 2340, 2170, 1710, 1605, 1610, 1540,
%1375, 1330, 1300, 1260, 1220, 1010, 890, and 805 $\rm
%cm^{-1}$ as well as a broad structure between 1000 and 600 $\rm cm^{-1}$.  
Features at 1605, 1375, 1300, and 1010 $\rm cm^{-1}$
show that some $\rm HCO_3^-$ sublimes in this temperature
interval. However, the weakness of these bands shows that the
contribution of bicarbonate to the total negative charge in the 200 -
240 K fraction is minor. The band at 1670 $\rm cm^{-1}$ and the broad
structure between 1000 and 600 $\rm cm^{-1}$ likely correspond to a small 
quantity of $\rm H_2O$ which is
embedded in the more refractory material (Schutte \& Buys 1961; Ryskin
1974). Likewise, the 2340 $\rm cm^{-1}$ feature can be ascribed to a
small amount of embedded $\rm CO_2$.  Like for the photolysed 
$\rm NH_3$/$\rm O_2$, the 1220 $\rm cm^{-1}$ may be due to the 
$\rm NO_2^-$ ion (Fig.~\ref{comp2res}). However,
the intensity of this band shows that $\rm NO_2^-$ accounts for only
$\sim$ half of the countercharge of the $\rm NH_4^+$ which sublimes in
the 200 - 240 K interval. The 2170 $\rm cm^{-1}$ feature indicates the
production of some $\rm OCN^-$ (Hudson et al. 2001), however, it is
too weak to make a substantial contribution to the countercharge.  We
conclude that an important fraction of this relatively volatile
residue component must consist of unknown anions. 
%Corresponding
%features that at 1710, 1540, 1330, 890 and 805
%$\rm cm^{-1}$. 
No sign of such species, or of the acids from which
they originate, was found in experiments with less than four initial
ice components (i.e., $\rm H_2O$/$\rm CO_2$/$\rm O_2$,
$\rm NH_3$/$\rm O_2$; sect. 3.1).  
It therefore seems probable that they are relatively
complex ions whose formation pathway involves all four original
components.

   \begin{figure}
   \centering
   \includegraphics[angle=-90,width=8.8cm]{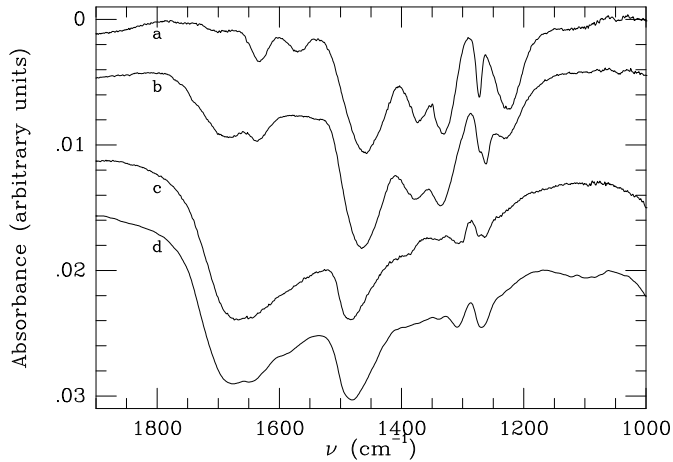}
\caption{Spectra of photolysed ice samples, after warm-up to 160 K,
enlargement of the 1900 - 1000 $\rm cm^{-1}$ region: a. 
$\rm NH_3$/$\rm O_2$ = 1/1 (expt. 7); b. 
$\rm H_2O$/$\rm CO_2$/$\rm NH_3$/$\rm O_2$=
1/0.5/0.7/1 (expt. 2); c. 10/2/1/1.1 (expt. 4); d. and 10/1.2/0.9/0.9 (expt. 5).
The spectra have been offset for clarity.
              }
         \label{comppartcomp}
   \end{figure}

As outlined in Sect. 1, a prime issue for the viability of the $\rm NH_4^+$
assignment of the 6.85 $\mu$m band concerns whether it is possible to
produce a clear $\nu_4$ spectral signature while features of counterions
are weak or absent as for the interstellar spectra. From
Fig.~\ref{comp4part} it can be seen that intense broad bands near 1320
and 1220 $\rm cm^{-1}$ are present once the ices have sublimed at
$\sim$ 180 K. As discussed above, these bands are caused by negative
ions, probably $\rm HCO_3^-$, $\rm NO_3^-$, and $\rm NO_2^-$.  Before
ice sublimation, these bands are however quite inconspicuous,
resulting in a smooth sloping spectrum in the 1400 - 1200 $\rm
cm^{-1}$ range with only minor substructure due to the 1300 and 1270
$\rm cm^{-1}$ features of unidentified photoproducts. 

To further investigate the influence of the $\rm H_2O$ matrix on the
appearance of the anion bands, Fig.~\ref{comppartcomp} compares the
1900 - 1000 $\rm cm^{-1}$ spectra of the ice samples $\rm H_2O$/$\rm
CO_2$/$\rm NH_3$/$\rm O_2$ = 0/0/1/1, 1/0.5/0.7/1, 10/2/1/1.1 and
10/1.2/0.9/0.9 (Expts. 2, 7, 4 and 5) All spectra were taken after
warm-up to 160 K and sublimation of $\rm NH_3$, which enables a better
view of the features of the photoproducts. Samples 2 and 7 show strong
bands due to negative ions at 1385 and 1335 $\rm cm^{-1}$ ($\rm
NO_3^-$) and at 1230 $\rm cm^{-1}$ ($\rm NO_2^-$). There is hardly a
sign of these or other features of anions in the $\rm H_2O$ dominated 
samples (expts. 4 and
5), even though the intensity of the $\nu_4$ $\rm NH_4^+$ band at
$\sim$ 1475 $\rm cm^{-1}$ is similar. This again shows that in the
$\rm H_2O$ dominated samples features due to anions are inconspicuous.

%What could explain the conspicuous absence of the cation spectral features
%in the $\rm H_2O$ dominated matrices ? One possibility would be that
%the bands of the negative ions are strongly suppressed, either by
%extensive broadening or by a substantial decrease in the band strength.
%The other possibility is that as long as $\rm H_2O$ is present cations
%other 

%Apparently the
%strength and/or profile of the features of the negative ions is
%strongly influenced by the $\rm H_2O$-dominated ice matrix.  
%This clearly demonstrates that in a water-dominated matrix the bands
%of the negative ions are strongly suppressed.

\subsection{Summary}

The photolysis of $\rm H_2O$/$\rm CO_2$/$\rm NH_3$/$\rm O_2$ ice
mixtures efficiently produces $\rm NH_4^+$. These experiments revealed 
the following four important spectral properties of $\rm NH_4^+$ and the
negative counterions which pose important constraints for an identification
of $\rm NH_4^+$ in interstellar ices:

1. During warm-up from 12 to 280 K, the $\rm
NH_4^+$ $\nu_4$ feature shows a pronounced shift from $\sim$1480 $\rm
cm^{-1}$ to $\sim$1420 $\rm cm^{-1}$. 

2. Between 12 to 120 K the $\rm NH_4^+$ $\nu_4$ feature grows smoothly by a 
factor 2.  

3. In $\rm H_2O$ ice three relatively weak features of $\rm NH_4^+$
are only evident after the ice sublimation at $\sim$ 180 K. These are
the $2\nu_2$ band near 3200 $\rm cm^{-1}$, the $\nu_1$+$\nu_5$
band near 3060 $\rm cm^{-1}$, and the $2\nu_4$ band near 2860 $\rm
cm^{-1}$.  

4. In $\rm H_2O$ dominated ice, the spectral features
of the {\it counterions} are only apparent after the sublimation of the ice, 
but are unconspicuous in the $\rm H_2O$ dominated ice
matrix. 

Concerning this last point, analysis of the spectrum of the residue 
shows that $\rm
HCO_3^-$, $\rm NO_3^-$, $\rm NO_2^-$ are probably present. The total
abundance of these species is somewhat unsufficient for balancing the
$\rm NH_4^+$, indicating the production of additional, yet
unidentified, anions.

The identifications of anions from the IR spectra are in some cases
tentative, while in other cases no identification was possible at
all. A positive identification of the anions and corresponding acid neutrals
may help to constrain the composition of interstellar ices (see Sect.
5 below). In future, the nature of the negative species could be studied
by analysis of the molecules that evaporate during the warm-up, for
example, by mass spectroscopy.

\subsection{Strength of the $\nu_4$ $\rm NH_4^+$ band}

\begin{table}
\caption[]{Band strength of the $\nu_4$ $\rm NH_4^+$ band}
\begin{flushleft}
\label{bandstrength}
\begin{tabular}{rrr}
\hline
\multicolumn{1}{c}{mixture} & \multicolumn{1}{c}{T (K)}
& \multicolumn{1}{c}{A (cm ion$^{-1}$)}\\
\hline
$\rm H_2O$/$\rm NH_3$/HCOOH\\
100/3.6/3.6 & 120 K & 4.4 (-17)\\
$\rm NH_3$/HCOOH=4/10 & 80 K & 4.0 (-17)\\
$\rm H_2O$/$\rm CO_2$/$\rm NH_3$/$\rm O_2$\\
10/1.2/0.9/0.9, + h$\nu$ & 150 K & $\geq$2.7 (-17)\\
\hline
\end{tabular}
\end{flushleft}
\end{table}

To obtain column densities, it is necessary to measure the intrinsic
strength of the $\rm NH_4^+$ $\nu_4$ feature. This can be derived from
experiments in which $\rm NH_4^+$ is formed by simple warm-up (no
photolysis) of an ice containing $\rm NH_3$ and an acid. In this case
the quantity of $\rm NH_4^+$ formed during warm-up is equal to the
amount of $\rm NH_3$ which is lost. Since the band strength of the
$\rm NH_3$ umbrella mode in various matrices is known (Kerkhof et
al. 1999), this allows a precise determination of the $\rm NH_4^+$
band strength. We performed two experiments involving $\rm NH_3$ with
formic acid (HCOOH; see Schutte et al. 1999 for experimental details).
The results are listed in Table~\ref{bandstrength}. Since the strength
may depend on the composition of the matrix and nature of the
counterion, the band strength of $\rm NH_4^+$ in the photolysis experiments 
may differ from the HCOOH/$\rm NH_3$
mixtures. No direct measurement in the photolysis experiments is
possible, because, besides acid-base reactions, other processes may
contribute to the $\rm NH_3$ destruction during the photolysis.
Therefore the amount of $\rm NH_3$ destroyed exceeds the formation of
$\rm NH_4^+$ by an unknown factor and only a lower limit to the $\rm
NH_4^+$ band strength can be obtained. The most constraining lower
limits are derived for the experiments with $\rm H_2O$-dominated
ices (see Table~\ref{bandstrength}). This is not surprising, since 
in these experiments features of
N-containing photoproducts other than $\rm NH_4^+$ are small
(Sect. 3.1).

We will adopt a standard band strength 
A($\nu_4$,$\rm NH_4^+$) = 4.4$\times$10$^{-17}$ cm molec.$^{-1}$.
This is close to the value of (2.5 - 3.5)$\times$10$^{-17}$ cm molec.$^{-1}$
found in aqueous solution (Lowenthal \& Khanna, in preparation).
As discussed in Sect. 3.1, the band strength varies by a factor 2
with temperature in our photolysis experiments.
It is unclear whether the standard band strength corresponds to the
highest or lowest bandstrength in the photolysis experiments.
We will therefore adopt a standard uncertainty of a factor 2 in this
value.

\section{Comparison to observations of Young Stellar Objects}

\begin{table*}
\caption[]{Properties of the 3.26, 3.48, and 6.85 $\mu$m features towards
YSO's.}
\begin{flushleft}
\label{featuresobs}
\begin{tabular}{lccccccccc}
\hline
\multicolumn{1}{c}{Object} & \multicolumn{1}{c}{N($\rm H_2O$)$^a$} & 
\multicolumn{1}{c}{$\rm {{F(45)}\over{F(100)}}$$^b$}
& \multicolumn{2}{c}{3.26 $\mu$m$^c$} & \multicolumn{1}{c}{3.48 $\mu$m$^c$} &
\multicolumn{2}{c}{6.85 $\mu$m} & ${{\tau(3.26)}\over{\tau(6.85)}}^h$ &
$\rm {{N(NH_4^+)}\over{N(H_2O)}}^i$ \\ 
& \multicolumn{1}{c}{$\rm 10^{18}$$\rm cm^{-2}$} & 
& \multicolumn{1}{c}{$\nu$($\rm cm^{-1}$)} & \multicolumn{1}{c}{$\tau$} & 
\multicolumn{1}{c}{$\tau$} & 
\multicolumn{1}{c}{$\nu$($\rm cm^{-1}$)} & \multicolumn{1}{c}{$\tau$$^d$} 
& & \% \\
\hline
\\
W33A                & 12 & 1.27 &      & $<$0.4 & $>$0.29 
& 1471 & 1.1 & $<$0.33 & 11 \\
NGC7538:IRS9        & 7 & 2.07 &      &        & 0.13 (0.02)
& 1473 & 0.29 & & 5 \\
GL2136           & 5.0 & 2.75 &      & $<$0.14 & 0.14 (0.02)    
& 1459 & 0.23 & $<$ 0.6 & 6 \\
S140:IRS1 & 2.8 & 3.24 & 3072 & 0.036 (0.007) & 0.027 (0.007)  
& 1451 & 0.16 & 0.23 (0.06) & 7 \\
          &     &  & 3085$^g$ & 0.050$^g$ & 0.050$^g$ & & & {\bf 0.31} \\
W3:IRS5             & 5.8 & 4.12 &      &        & 0.13    
& 1459 & 0.25 & & 5 \\
MonR2:IRS3          & 1.6$^f$ & 4.32 & 3071 & 0.049 (0.007) & 
0.036   & 1437 & 0.23 & 0.21 (0.03) & 17 \\
AFGL7009S           & 12 &     &     &      &        &         
1462 & 1.1 & & 10 \\
GL989            & 3 &     & 3098$^g$ & 0.069$^g$ & 0.069$^g$ & 1471 
& 0.11 & {\bf 0.63} & 4 \\
$\rho$ Oph Elias 29 & 3 &    &      & $<$0.03 & 0.089 (0.007) & 1470 
& 0.07 & $<$0.4 & 3 \\
                    &   &    & 3080$^g$ & 0.085$^g$ & 0.065$^g$ & & 
& {\bf 1.2} \\
HH100     & 2.4 &   & 3087 & 0.032 (0.010) & 0.041 (0.005)
& 1470 & 0.15 (0.06) & 0.21 (0.09) & 8 (3) \\
NGC7538:IRS1 & 3 &     & 3065 & 0.078 (0.013) & 0.052 (0.014)  
& 1455 & 0.32$^e$ & 0.24 (0.04) & 13 \\
\hline
\end{tabular}
\end{flushleft}
\begin{list}{}{}
\item[$^a$] Water column density obtained from the integrated
depth of the 3 $\mu$m band: NGC7538:IRS9/Allamandola et al. 1992;
AFGL7009S/Dartois et al. 1999a (from 6 $\mu$m feature); 
W33A/Gibb et al. 2000 (obtained by fitting the 3 $\mu$m
feature with laboratory analog spectra); AFGL 989/Smith et al. 1989;
AFGL 2136/Schutte et al. 1996a; Elias 29/Boogert et al. 2000; 
S140:IRS1(or GL2884)/Willner et al. 1982; W3:IRS5/Allamandola et al. 1992;
MonR2:IRS3/Smith et al. 1989; HH100(or RCra:IRS1)/Whittet et al. 1996; 
NGC7538:IRS1/Willner et al. 1982.
\item[$^b$] Ratio of flux at 45 and 100 $\mu$m; from Keane et al. 2001
\item[$^c$] Ground-based observations (Sellgren et al. 1995, 
Brooke et al. 1996, Brooke et al. 1999), unless otherwise noted. Error noted
in parentheses.
\item[$^d$] Keane et al. 2001; error typically 20 \%, unless otherwise noted
\item[$^e$] From the ISO-SWS database
\item[$^f$] It is unclear whether the 3 $\mu$m feature of MonR2:IRS3 is
fully caused by $\rm H_2O$ ice, due to the large width of this band 
(Smith et al. 1989). Therefore this number should be considered an upper limit.
\item[$^g$] Observed by ISO-SWS (Bregman et al. 2001)
\item[$^h$] Boldface entries give the 3.26 $\mu$m depth
derived from ISO-SWS data (Bregman et al. 2001). Errors are given in
parentheses.
\item[$^i$] The uncertainty, due to the error in the bandstrength, is
a factor $\sim$ 2.
\end{list}
\end{table*}

In this section we investigate a number of criteria for the identification
of the interstellar 6.85 $\mu$m absorption with the $\nu_4$ feature of the
ammonium ($\rm NH_4^+$) ion. First of all, the feature should provide a good
match to the observed band. Second, the characteristic redshift of
the feature with temperature (Sect. 3.1; Fig.~\ref{redshift}) should show 
up in the observational data.  Third, additional bands of
$\rm NH_4^+$ are sought, specifically the
$2\nu_2$, $\nu_1+\nu_5$, and $2\nu_4$ features near 3200, 3060, and 2860
$\rm cm^{-1}$ (3.12, 3.27, and 3.50 $\mu$m). Fourth, there should be
no spectral structure due to counter-ions that is inconsistent with the
interstellar data. It will be shown that the
available observational information indicates that all these criteria
are satisfied.  For further reference, Table~\ref{featuresobs} 
lists the relevant
observational data for all lines of sight where high quality
spectra of the 6.85 $\mu$m feature (i.e., by ISO-SWS) are available.
Data on the interstellar 3.26 and 3.48 $\mu$m features are included
because they are possibly associated with the $\rm NH_4^+$ $\nu_1+\nu_5$
and $2\nu_4$ bands (see below). The depth
of the 6.85 $\mu$m band was measured from the original data. These were
obtained from the ISO-SWS database (for NGC7538:IRS1) and from
Keane et al. 2001 (all other objects). The depth was obtained by
subtracting a linear baseline in the log(F) vs $\mu$m plane.
The baseline was drawn through the 5.5 and 7.5 $\mu$m points.

\subsection{Comparison with the interstellar 6.85 $\mu$m feature}

\subsubsection{W33A}

   \begin{figure}
   \centering
   \includegraphics[width=9.4cm]{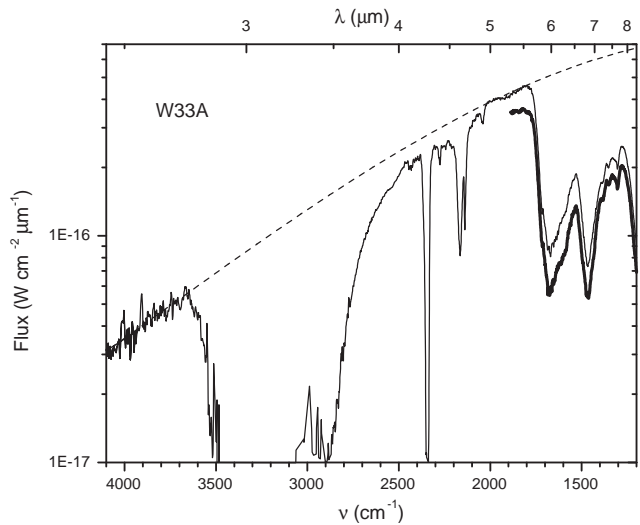}
\caption{ISO/SWS 1 (thin line) and SWS 6 spectra (1900 - 1200 $\rm cm^{-1}$
thick line) of W33A
(from Gibb et al. 2000 and Keane et al. 2001, respectively).
The smooth dashed curve is a third-order polynomial continuum fit.
              }
         \label{W33A}
   \end{figure}

   \begin{figure}
   \centering
   \includegraphics[angle=-90,width=8.8cm]{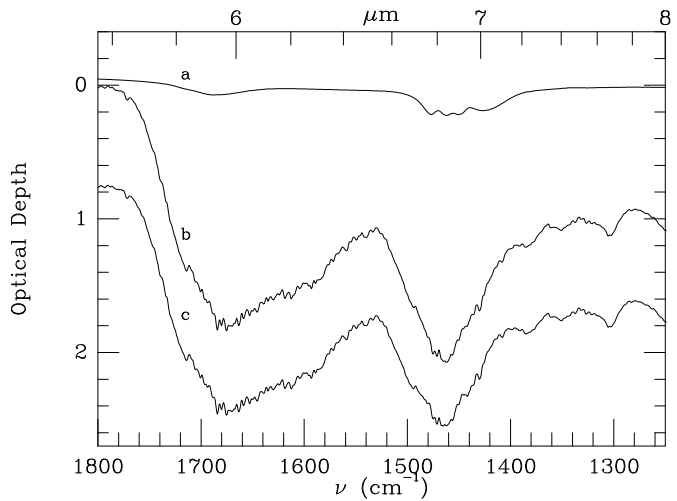}
\caption{Correction of the 6.85 $\mu$m band of W33A for the contribution
of $\rm CH_3OH$. a. $\rm H_2O$/$\rm CH_3OH$/$\rm CO_2$ = 1/1/1, at 112 K
(from Ehrenfreund et al. 1999; band strength has been scaled to correspond
to the $\rm CH_3OH$ column density, see text); 
b. Optical depth spectrum of W33A obtained
by subtraction of the continuum from the SWS 6 data (See Fig.~\ref{W33A}); 
c. Corrected optical depth spectrum of W33A obtained by subtraction of curve 
a from curve b.  The vertical scale is identical for all
curves, however, curve c has been offset for clarity.
              }
         \label{W33Acorr}
\end{figure}

   \begin{figure}
   \centering
   \includegraphics[width=9.4cm]{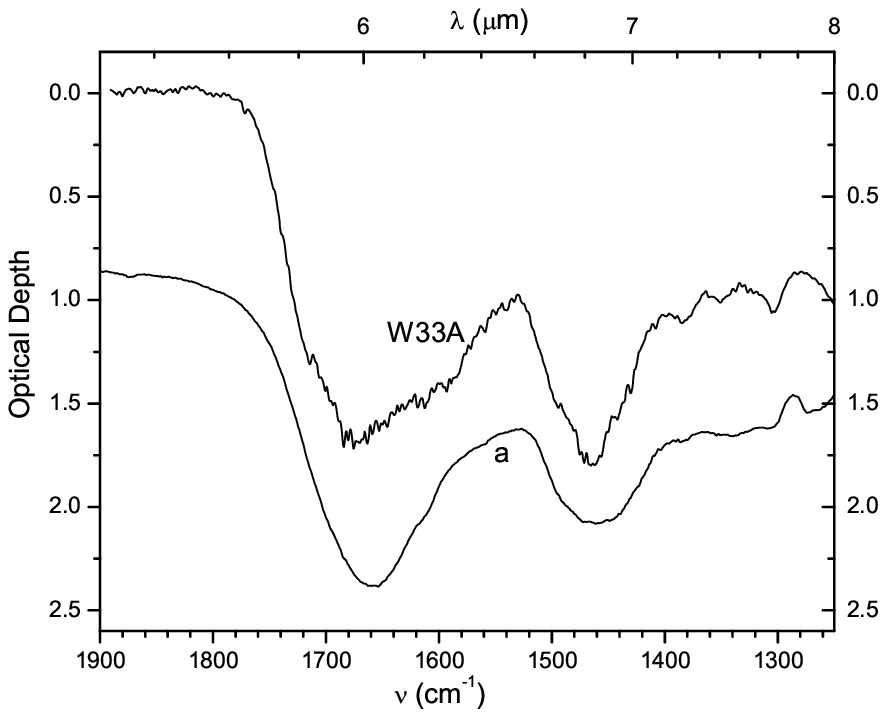}
\caption{The optical depth spectrum of W33A after correction
for $\rm CH_3OH$ (Fig.~\ref{W33Acorr}), compared to: a. Photolysed
$\rm H_2O$/$\rm CO_2$/$\rm NH_3$/$\rm O_2$ = 10/2/1.1/1 (expt. 4),
average of 12 K, 120 K, 180 K (x2) spectra. Curve a has 
been offset for clarity.
              }
         \label{W33Acomp}
\end{figure}

Fig.~\ref{W33A} shows the 4000 - 1200 $\rm cm^{-1}$ (2.5 - 8.3 $\mu$m)
spectra of the high mass YSO W33A obtained by ISO/SWS in observation modes 1 
and 6 (de Graauw
et al. 1996). Spectra were adopted from Gibb et al. (2000) and Keane et
al. (2001), respectively. While the SWS6 spectrum gives the highest
S/N, the wide range of the SWS1 spectrum allows an accurate baseline
determination. We defined the baseline over the entire 4000 -
1200 $\rm cm^{-1}$ range by drawing a smooth third
order polynomial through the continuum regions 4100 - 3700 $\rm
cm^{-1}$, 2010 - 1980 $\rm cm^{-1}$, and 1840 - 1810 $\rm cm^{-1}$ 
(Fig.~\ref{W33A}). This procedure differs in some
respects from the baseline definition in Keane et al. and Gibb et
al. Keane et al. defined the baseline only over a limited data range
(1900 - 1300 $\rm cm^{-1}$), and included the region around 1300 $\rm
cm^{-1}$, just before the onset of the silicate feature, in the baseline fit.
It is however clear from Fig.~\ref{W33A} that, at least
for W33A, there is considerable absorption in this region. Gibb et
al., while also fitting the baseline over a similar broad range,
included the 2500 - 2440 $\rm cm^{-1}$ region in the fitting
procedure. However, it is probable that both the long-wavelength
shoulder of the $\rm H_2O$ band (Willner et al. 1982; Smith et
al. 1989, see also Fig.~\ref{W33A}) as well as the broad $\rm H_2O$
combination feature near 2220 $\rm cm^{-1}$ (Gibb et al. 2000) extend into this
region. Indeed, by excluding it we obtain a smoother baseline
than Gibb et al. (i.e., third, rather than fourth order polynomial).

Although there is a slight offset between the SWS1 and SWS6 data,
their overall spectral shape is quite consistent.  Therefore the SWS6
data were converted to an optical depth scale by subtraction of the
baseline of Fig.~\ref{W33A}. The result is shown in
Fig.~\ref{W33Acorr}.  The small offset between the SWS1 and SWS6
data was compensated by normalizing to 0 in the 1900 - 1800 $\rm
cm^{-1}$ region.

The CH$_3$ deformation feature of methanol falls near 6.85 $\mu$m and
will contribute a fraction of the interstellar band. 
The methanol column density towards W33A, as determined
from the $\nu_3$ and combination bands near 3.54 and 3.9 $\mu$m, respectively,
equals 1.85 $\times$ 10$^{18}$ $\rm cm^{-2}$ (Dartois et al. 1999b). 
Fig.~\ref{W33Acorr} compares the optical depth spectrum of W33A with 
the $\nu_3$ band of methanol scaled to this column density. This feature
was measured in a mixture $\rm H_2O$/$\rm CH_3OH$/$\rm CO_2$ = 1/1/1
(Ehrenfreund et al. 1999). This matrix should be
representative of the ices in which methanol in circumstellar
objects is embedded (Gerakines et al. 1999; Dartois et al. 1999a). 
It is clear from Fig.~\ref{W33Acorr} that
methanol only gives a small contribution to the observed 6.85 $\mu$m
band of $\sim$ 20 \% (see also Grim et al. 1991; Schutte et al. 1996).
To enable a more
suitable comparison with the laboratory $\rm NH_4^+$ spectra, we have
subtracted this small contribution from the W33A spectrum. The resulting curve
is shown in Fig.~\ref{W33Acorr}. Since the correction is small, 
using spectra of methanol in other matrices, such as pure methanol, 
or a water-dominated ice, gives a very similar result.

Fig.~\ref{W33Acomp} compares the 6.85 $\mu$m
band of W33A (after subtraction of the methanol feature) 
with the $\nu_4$ band of $\rm NH_4^+$ as produced
in one of our water-dominated samples after photolysis and warm-up
(expt. 4). To take into account the variety of dust temperatures
probed by the line of sight, the laboratory spectrum is an average of the
12, 120, and 180 K (x2) spectra. We note that good overall matches 
can be obtained with all photolysed water-dominated ices (expts. 3, 4 and 5), 
since they produce little substructure in the 1400 - 1250 $\rm cm^{-1}$
(7.1 - 8.0 $\mu$m) range. On the other hand, the photolysed ice mixtures 
with less water show strong peaks in this region due to $\rm NO_3^-$
and $\rm NO_2^-$ (Sect. 3.1; Fig.~\ref{comppartcomp}), which have no 
interstellar counterparts.

Due to the superposition of ices at different temperatures along the
line of sight, components of different volatility will have different
average temperature. This is illustrated by the temperature averaged
laboratory match of the
W33A spectrum (Fig.~\ref{W33Acomp}). The $\rm NH_4^+$ band arises 
primarily in the 120 and 180 K components, since at 12 K the intensity 
of this feature is
small (Fig.~\ref{comp4part}). On the other hand, the $\rm H_2O$ ice features
are produced in the 12 and 120 K spectra, since at T $\geq$ 170 K 
$\rm H_2O$ sublimes. Thus, the average temperature
of the solid $\rm H_2O$ in the laboratory spectrum is 66 K, while the average 
temperature of $\rm NH_4^+$
is $\sim$ 150 K. This clearly shows that temperatures derived by fitting the 
6.85 $\mu$m band should not be taken as representative of the temperature
of more volatile components located in cooler regions along the line of
sight.

\subsubsection{MonR2:IRS3}

   \begin{figure}
   \centering
   \includegraphics[width=9.4cm]{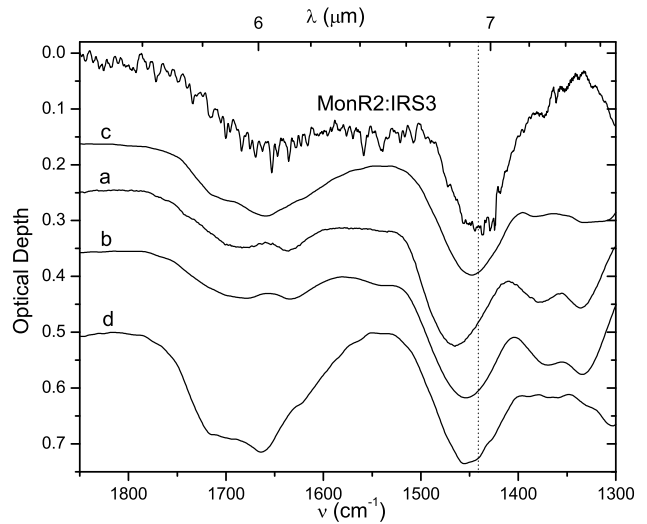}
\caption{The optical depth spectrum of MonR2:IRS3 (From ISO/SWS; Keane et
al. 2001) compared with some photolysed laboratory samples:
a. $\rm H_2O$/$\rm CO_2$/$\rm NH_3$/$\rm O_2$ = 1/0.5/0.7/1 at 160 K (expt.2);
b.same, at 170 K; c. 10/2/1.1/1 at 180 K (expt. 4); d. 10/1.2/0.9/0.9 at 180 K
(expt. 5). The laboratory spectra are offset for clarity.
              }
         \label{MonR2comp}
\end{figure}

MonR2:IRS3 is an extreme object in a number of ways.  First, the solid
$\rm CO_2$ abundance is exceptionally low ($\sim$ 1 \%; Keane et
al. 2001).  Second, its 6.85 $\mu$m feature is the reddest of all
YSO's (Table~\ref{featuresobs}). Third, the strength of the 6.85
$\mu$m feature relative to the column density of $\rm H_2O$ is very
high (Table~\ref{featuresobs}). This all may indicate that the ice
along the line of sight is exceptionally warm, as is also indicated by
the high flux ratio F(45 $\mu$m)/F(100 $\mu$m) (Keane et al.  2001;
Table~\ref{featuresobs}).  Indeed, the large relative intensity of the
6.85 $\mu$m band, assuming that the carrier of this band is a
relatively refractory species, suggests that $\rm H_2O$ ice has begun
to evaporate along most of the line-of-sight. This is confirmed by the
large column density of water vapor (Boonman et al. 2000).  Ice
sublimation should greatly enhance the visibility of features of more
refractory components, making MonR2:IRS3 an excellent testing ground
for the presence of $\rm NH_4^+$ and associated counterions.

Fig.~\ref{MonR2comp} compares the 6.85 $\mu$m band of MonR2:IRS3
with the $\nu_4$ $\rm NH_4^+$ features obtained in several of our
experiments.  This set of experimental data will further down be used 
to evaluate the presence of other interstellar $\rm NH_4^+$ bands. No
correction was made for the contribution of methanol to the 6.85
$\mu$m band, since no $\rm CH_3OH$ has been observed towards this
source (abundance $\leq$ 4.5 \%; Dartois et al. 1999b). The best
matches are obtained for the photolysed water-rich ices (expts. 4 and 5), 
since the water-poor ice (expt. 2) gives strong
features at 1385 and 1335 $\rm cm^{-1}$ due to $\rm NO_3^-$ which are
not observed. In all cases, the position of the laboratory band is
somewhat blue-shifted from the MonR2:IRS3 feature. This situation
could be mended by selecting laboratory spectra at higher
temperatures. However, for such high temperatures a broad band near 
1320 $\rm cm^{-1}$, which is caused by a superposition of features
of various negative ions (Fig.~\ref{comp4part}, Sect. 3.1) shows up strongly,
resulting in a decrease of the overall quality of the
match. Even at the temperatures selected for Fig.~\ref{MonR2comp}  too much
absorption appears to be present in the 1350 $\rm cm^{-1}$
region. However, this difference could be caused by
the original choice of baseline, which made the a priori assumption of
little absorption in the 1350 $\rm cm^{-1}$ region (Keane et al. 2001).  
As for W33A, considerable absorption may be present in this region.
Thus, the position of the 6.85 $\mu$m feature of
MonR2:IRS3 is not fully reproduced by the experimental spectra,
although the difference is small ($\sim$ 3-10 $\rm cm^{-1}$) for
the three $\rm H_2O$-dominated samples. In view of the strong
matrix dependence of the position of this band (Fig.~\ref{redshift}), it seems 
plausible that the offset may be caused by
a shift associated with the composition of the interstellar
matrix.

Besides the 6.85 $\mu$m band, the 6 $\mu$m feature of MonR2:IRS3 also
has a counterpart in the laboratory spectra. Especially
expt. 4 at 180 K gives a band which is quite similar (Fig.~\ref{MonR2comp}).
At this temperature
most $\rm H_2O$ has sublimed, and the laboratory features near 6 $\mu$m 
are dominated by a superposition of bands from various negative ions,
i.e., $\rm HCO_3^-$ and one or more unidentified species (Fig.~\ref{comp2res}; 
Sect. 3.1).
This suggests that most of the 6 $\mu$m feature of MonR2:IRS3
is caused by non-volatile components rather than $\rm H_2O$ ice. Such a
conclusion is consistent with the observations, which show that the
6 $\mu$m feature of MonR2:IRS3 is 3.5 times stronger than expected at the
$\rm H_2O$ column density derived from the 3 $\mu$m band (Keane et al. 2001).

\subsection{The redshift with temperature}

   \begin{figure}
   \centering
   \includegraphics[angle=-90,width=8.8cm]{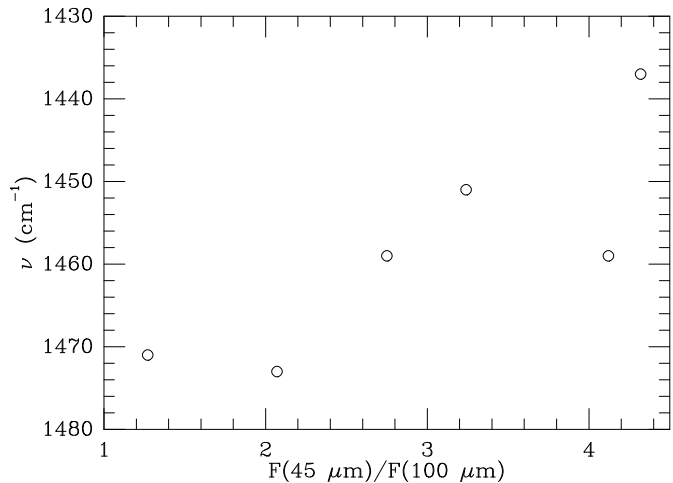}
\caption{Position of the 6.85 $\mu$m feature vs. F(45 $\mu$m)/F(100 $\mu$m)
for the YSO's listed in Table~\ref{featuresobs}.
              }
         \label{redshiftobs}
\end{figure}

It was shown in Sect. 3.1 that the ammonium band shifts strongly
towards the red with increasing temperature. Even though the position of
the $\rm NH_4^+$ band depends considerably on the starting matrix, this 
general trend held for all our experiments. Thus, if the
ammonium ion is the carrier of the interstellar 6.85 $\mu$m feature,
a correlation should be found between the position of the 6.85 $\mu$m
feature and indicators of the dust temperature along the line
of sight.

Fig.~\ref{redshiftobs} plots the position of the 6.85 $\mu$m band in the 
data set of Keane et al. (2001) against the ratio of the fluxes at 45 and 100
$\mu$m as measured by ISO/LWS. This ratio is a general indicator
of the dust temperature along the line of sight. It can be seen that there is a
clear correlation between the position of the band and the flux ratio,
in the sense that the interstellar band shows a strong redshift
with increasing average dust temperature. A similar conclusion was already
formulated by Keane et al., with the difference
that they interpreted
the change as due to a systematic variation with line of sight
temperature in the contribution
of two independent components of the 6.85 $\mu$m feature. It must
be stressed that interpretation of the temperature dependence of the
band as either a variation of 
independent components, or a shift of a single component,
is solely a matter of interpretation and cannot be decided on the
basis of the interstellar band profile, since the feature does
not show substructure.

The systematic redshift of the 6.85 $\mu$m band with line-of-sight dust
temperature is in good agreement with the
trend observed for the $\nu_4$ feature of the ammonium ion. Comparing
Fig.~\ref{redshiftobs} and Fig.~\ref{redshift} of Sect. 3.2.1, it can 
be seen that the spectral
region in which the interstellar feature is found falls well within
the range spanned by the position of the $\nu_4$ band. Thus,
the systematic redshift of the interstellar 
6.85 $\mu$m band gives strong support 
to its identification with the $\nu_4$ mode of $\rm NH_4^+$.

The range of
positions of the interstellar band corresponds to laboratory
temperatures of $\sim$ 120 - 240 K. It must be noted though that 
the position of the feature not only depends on temperature,
but also on the ice matrix (Fig.~\ref{redshift}). Therefore, this 
temperature range 
should be taken as a rough indication.
Nevertheless, it seems clear that solid $\rm NH_4^+$ 
is found at higher temperatures than more volatile ice components.
Matching the profile of the 3 $\mu$m $\rm H_2O$ feature typically
yields 20 - 80 K (Smith et al. 1989), while matching the profile
of the 15.2 $\mu$m bending mode of $\rm CO_2$ gives 110 - 140 K
(Gerakines et al.  1999). This difference in temperature is
consistent with the refractory nature of $\rm NH_4^+$, which will
survive in the warmer regions close to the embedded source where the
more volatile ice components have sublimed. In addition, the band strength
of $\rm NH_4^+$ increases considerably above $\sim$ 100 K (Sect. 3.1;
Fig.~\ref{comp4part}). This effect will 
minimize the contribution of the coldest regions to the 6.85 $\mu$m band.

\subsection{Additional $\rm NH_4^+$ features}

   \begin{figure*}
   \centering
   \includegraphics[width=16cm]{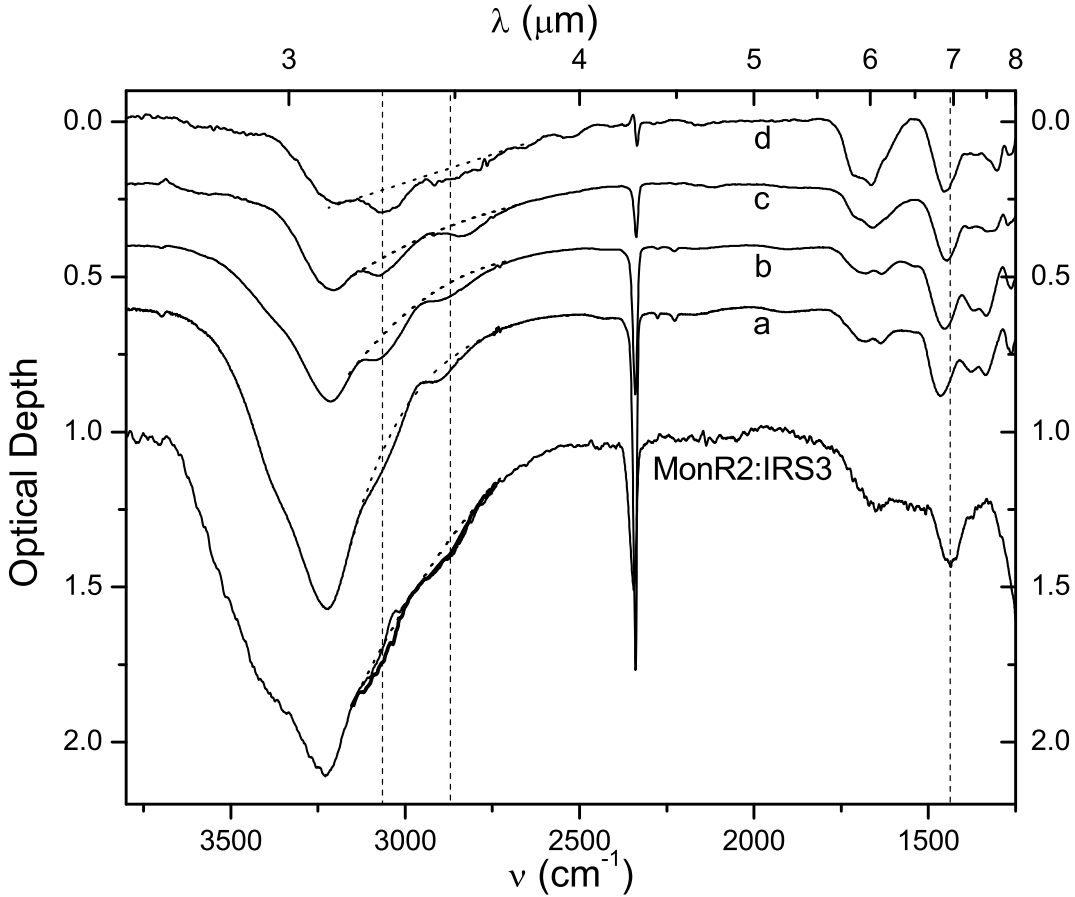}
\vskip-30pt
\caption{Comparison between the 3.26, 3.48 and 6.85 $\mu$m
features of MonR2:IRS3 and the $\nu_2+\nu_4$, $2\nu_4$ and $\nu_4$
features of $\rm NH_4^+$ in a number of laboratory samples. The
various $\rm NH_4^+$ features are indicated by vertical dashed lines. 
The labelling of the laboratory
spectra corresponds to the description given with
Fig.~\ref{MonR2comp}.  Dotted lines indicate the adopted baselines for
the two features.  The observational data comprise
ground-based observations between 3.16 - 3.65 $\mu$m (Sellgren et
al. 1995; bold solid line) and ISO/SWS data (2.6 -
8 $\mu$m; Gibb et al. 2001; solid line). The MonR2:IRS3 data were
converted to optical depth by subtraction of a 2$^{nd}$ order
polynomial baseline defined by fitting the continuum regions around
3800 and 2000 $\rm cm^{-1}$ in the original flux data. All spectra
were smoothed to resolution 400. The spectra have been
offset for clarity.
              }
         \label{MonR2comp3p26}
\end{figure*}

   \begin{figure}
   \centering
   \includegraphics[width=9.4cm]{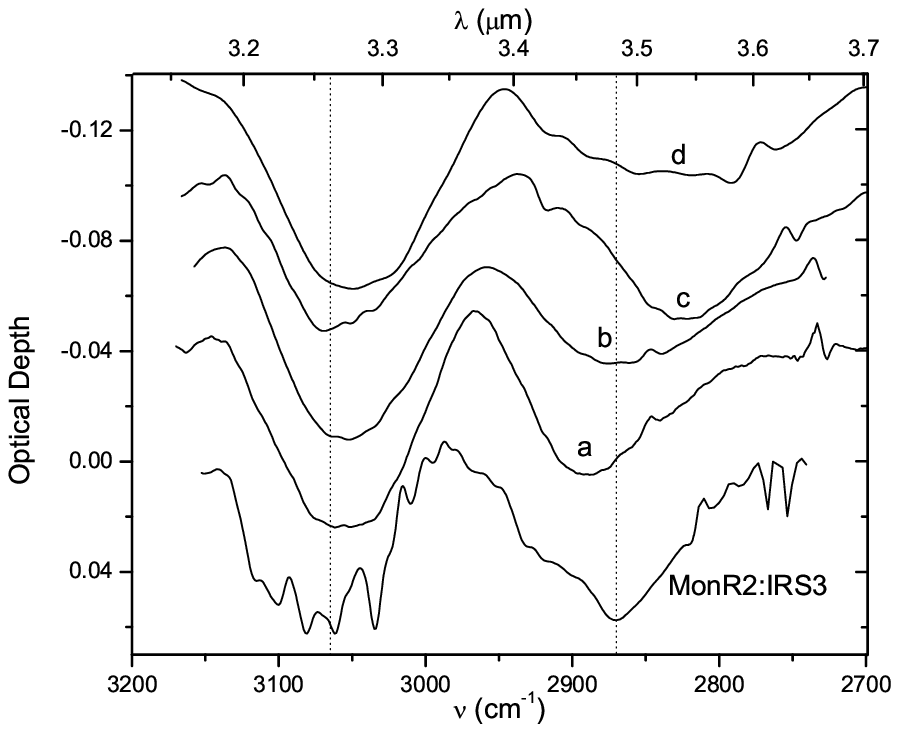}
\caption{Comparison between the 3.26 and 3.48 $\mu$m 
(3075 and 2875 $\rm cm^{-1}$) features
of MonR2:IRS3 and the $\nu_2+\nu_4$ and $2\nu_4$ bands of
$\rm NH_4^+$ in the laboratory samples of Fig.~\ref{MonR2comp3p26}, after
subtraction of the baselines indicated in this figure. The laboratory
spectra have been offset for clarity.
              }
         \label{MonR2comp3p26_2}
\end{figure}

Besides the $\nu_4$ feature near 6.85 $\mu$m, the ammonium ion shows a
number of additional features, i.e.,  $2\nu_2$, $\nu_1+\nu_5$,
and $2\nu_4$. These are centered near
$\sim$ 3200 $\rm cm^{-1}$ (3.12 $\mu$m), 
$\sim$ 3060 $\rm cm^{-1}$ (3.27 $\mu$m), and
$\sim$ 2860 $\rm cm^{-1}$ (3.50 $\mu$m), respectively (Figs.~\ref{comp4} \& 
~\ref{comp4res},
Sect. 3.2.1). 
The optical depth of each of these bands is typically 30 \% of that of the 
$\nu_4$ feature.

The $2\nu_2$ feature would be very difficult to distinguish,
since it nearly coincides with the peak of the 3 $\mu$m water band
(Smith et al. 1989).  The two other bands, while relatively weak, may
be observable as substructure on the red wing of the water band.
Indeed, features have been observed towards many protostellar sources
near these positions (e.g., Sellgren et al. 1995; Brooke et al. 1996;
Brooke et al. 1999; summary in Table~\ref{featuresobs}). For example,
Fig.~\ref{MonR2comp3p26} shows the mid-IR spectrum of MonR2:IRS3
obtained by ISO-SWS (Gibb et al. 2001) and by ground-based
spectroscopy (between 3.16 - 3.65 $\mu$m; Sellgren et al. 1995). As
already mentioned above, this source appears to be exceptionally warm and
therefore should have enhanced abundances of refractory components
like $\rm NH_4^+$.

To facilitate comparison with the laboratory data, a continuum has
been subtracted from the MonR2:IRS3 original flux data (Ground-based as
well as ISO-SWS). This continuum was
obtained using the ISO-SWS data by drawing a second order polynomial
through the zero absorption regions around 3800 and 2000 $\rm
cm^{-1}$. Before continuum subtraction, the ground-based data were 
multiplied by 0.88 to match them to the ISO-SWS spectrum. 

A small excess
is apparent in the ISO-SWS data relative to the ground based spectrum around 
3.29 $\mu$m. This excess is obviously caused by the presence of the
well-known emission feature in the wide aperture ISO data (This is a
common phenomenon for ISO observations towards YSO's; Gibb et al. 2001).

As already discussed by Sellgren et al., MonR2:IRS3 shows two features
at 3.26 and 3.48 $\mu$m. Fig.~\ref{MonR2comp3p26} also
shows the $\nu_2+\nu_4$ and $2\nu_4$ $\rm NH_4^+$ 
bands as obtained in our experiments. 
To allow a detailed comparison, we subtracted baselines from
both the laboratory and interstellar spectra, as indicated in Fig.
~\ref{MonR2comp3p26}. The resulting optical depth curves are shown in 
Fig.~\ref{MonR2comp3p26_2}. Only the ground-based observations were used,
due to the ``contamination'' of the ISO spectra with the 3.3 $\mu$m emission
feature.

The comparison in Figs.~\ref{MonR2comp3p26} and \ref{MonR2comp3p26_2} 
indicates a good general agreement between
the 3.26 and 3.48 $\mu$m features of MonR2:IRS3 and the 
 $\nu_2+\nu_4$ and $2\nu_4$ $\rm NH_4^+$ bands. Nevertheless, no single
spectrum is able to fit in detail both the position and relative
intensity of the observed features. However, the variation among the laboratory
spectra shows that the features are quite matrix and temperature
dependent, and it therefore seems plausible that this moderate difference
could derive from matrix effects. Besides the agreement in profile,
the agreement between the intensity of the laboratory and the
MonR2:IRS3 features lends support to an identification with
$\rm NH_4^+$.

It is clear from Fig.~\ref{MonR2comp3p26} that if the 3.26 and 3.48
$\mu$m features of MonR2:IRS3 are caused by $\rm NH_4^+$, the ion would
also partially account for the overall broad absorption in the 3.1 $\mu$m
feature. Indeed, some 50 \% of the unidentified excess absorption between 
3.25 and 3.8 $\mu$m (3080 - 2630 $\mu$m$^{-1}$),
which cannot be accounted for by $\rm H_2O$ ice (Smith et al. 1989),
could arise from the ion.

If the 3.26, 3.48, and 6.85 $\mu$m features are all due to the
same carrier, a tight correlation between these bands is expected.
Ground based observations of the 3.26 $\mu$m feature 
exist for four of the objects whose 6.85 $\mu$m feature has been observed by 
the SWS. 
(Table~\ref{featuresobs}).  For these objects the data give a stable 
ratio $\tau$(3.26)/$\tau$(6.85) = 0.22 $\pm$ 0.02. The constancy of the
ratio is even somewhat better than what could be expected on the basis
of the rather large uncertainty in the depth of the 3.26 $\mu$m band
(Table 4). Besides from telluric interference, this
error derives from the baseline definition, since this shallow feature is
located in a complex
spectral region, where various spectral features are located, such as
the broad feature of ammonia hydrate around 3.5 $\mu$m (2860 $\rm cm^{-1}$)
and the 3.36 $\mu$m 
(2980 $\rm cm^{-1}$) feature of $\rm CH_3OH$, as well as possibly
structure caused by scattering (Dartois \& d'Hendecourt 2001; Brooke et al.
1999). These features may contribute to the broad long wavelength
shoulder of the 3 $\mu$m band, on which the feature is superimposed.

For some of these objects ISO-SWS observations on the 3.26 $\mu$m
feature have also been reported (Bregman et al. 2001). Using these
data, there is little or no correlation between the 3.26 and 6.85
$\mu$m bands (Table~\ref{featuresobs}).  Indeed, the ground based and
ISO data give conflicting results on the depth of the 3.26 $\mu$m
band, most notably in the case of $\rho$ Oph Elias 29, where the
discrepancy is at least a factor of 3. It not quite clear what causes
these differences and which dataset is most reliable. 
For Elias 29, Bregman et al. used a speed 3 scan in
AOT SWS1. However, a higher quality SWS6 scan in this region does not
show the 3.26 $\mu$m feature, consistent with the
ground-based observations (Boogert et al. 2000; Boogert, private 
communication). A general problem
with the ISO data could be their wide aperture. For this reason, the
3.3 $\mu$m emission feature often shows up in the SWS spectra of embedded
YSO's (Gibb et al. 2001; see also Fig.~\ref{MonR2comp3p26}). 
Indeed, some 3.3 $\mu$m
emission seems consistent with the ISO spectra of S140 and GL
989 (Gibb et al. 2001). The presence of even a small amount of this feature 
would severely
complicate a correct baseline definition and therefore introduce
a large error in the depth of the shallow 3.26 $\mu$m band. 
On the other hand, in
ground-based observations the center of the 3.26 $\mu$m band has an
enhanced noise level due to telluric methane absorption.  However, the
ISO and ground-based observations also deviate for the 3.48 $\mu$m
band (Table 4), even though it falls in a relatively undisturbed
region. This deviation therefore reinforces the concern
that errors in the baseline definition
caused by the 3.3 $\mu$m emission feature is a major pitfall when using
ISO-SWS data to study the 3.26 and 3.48 $\mu$m absorption bands.

Good candidates to search the 3.26 $\mu$m feature are NGC7538:IRS9
and W3:IRS5, for which we predict, from the strength of the 6.85 $\mu$m
band, $\tau$(3.26) $\approx$ 0.064 and 0.055, respectively. A detection
of these bands would be an important test of the $\rm NH_4^+$ assignment.

No tight correlation exists between the 3.26 $\mu$m and 3.48 $\mu$m
features. Relative intensities vary between 0.13 (MonR2:IRS3) and 1.3
(Elias 29; see Table~\ref{featuresobs}).  While the relative intensity
of the feature of MonR2:IRS3 is consistent with an assignment of the
3.48 $\mu$m band to $\rm NH_4^+$ (Fig.~\ref{MonR2comp3p26_2}), it is
clear that in general other species must also contribute to this
feature. This is not too surprising, since additional structure is
present in this region. In particular, the long wavelength shoulder or
wing of the 3 $\mu$m $\rm H_2O$ band, which extends from $\sim$ 3.1 -
3.8 $\mu$m (3200 - 2600 $\rm cm^{-1}$; Smith et al.  1989; their
Fig. 6) produces structure in this region (Smith et al. 1989; their
Fig. 4).  This absorption will blend in with any additional absorbers,
e.g., the CH stretching mode of aliphatic hydrocarbons and the $\sim$
3.5 $\mu$m feature of ammonia hydrate (e.g., Dartois \& d'Hendecourt
2001).

\subsection{The 3.26 $\mu$m feature and PAHs}

Previously, an assignment of the 3.26 $\mu$m feature to the CH
stretching modes of polycyclic aromatic hydrocarbons has been proposed
(PAHs; Sellgren et al. 1995; Bregman et al. 2000). Such an assignment
would of course limit the contribution of the $\nu_2+\nu_4$ $\rm
NH_4^+$ feature to this band (Sect. 4.3). Here we quantitatively
assess the contribution of the PAH CH stretching mode to the 3.26
$\mu$m feature of MonR2:IRS3.

In support of, at least, part of the 3.26 $\mu$m feature deriving from the
PAH CH tretching mode is
the observation of a weak absorption band at 11.2 $\mu$m towards
MonR2:IRS3, which is ascribed to the out-of-plane bending mode of
isolated CH groups on the aromatic skeleton (Bregman et al. 2000). The
intensity of this band can be used to constrain the PAH contribution
to the 3.26 $\mu$m feature. The integrated optical depth are 
$\tau_{int}$(11.2) = 0.81 cm$^{-1}$ and
$\tau_{int}$(3.26) = 5.6 cm$^{-1}$.  The ratio of the band strength of
the CH stretching mode to that of the CH bending mode for isolated CH
groups equals 0.5 for gaseous, neutral PAHs. For ionized or solid
state PAHs the ratio is at least 4 times smaller (Bregman et al. 2000
and references therein).  Therefore the CH groups associated with the
11.2 $\mu$m feature can, at most, cause an absorption of
$\tau_{int}$(3.26 $\mu$m;PAH) = 0.4 cm$^{-1}$, i.e., only 7 \% of the
observed intensity.  Besides isolated CH groups, aromatic species will
have CH groups on their periphery which have 1 or more adjacent CH
groups. For these the CH bending mode falls at longer wavelength (11.6
- 13.6 $\mu$m; Hony et al. 2001). Besides discrete features near 12.7
and 13.5 $\mu$m, the bending modes of these groups give rise to a
broad plateau between 11 - 13 $\mu$m caused by overlapping
absorptions. In emission, the fraction of the total emission
associated with the CH bending modes (i.e., 11.2, 12.7, 13.5 $\mu$m
bands and the plateau) which is emitted by the 11.2 $\mu$m band varies
between 0.3 - 0.7 (Hony et al. 2001). Since the band strength of the
CH bending mode for adjacent aromatic CH groups is two times weaker
than for isolated CH groups, this implies that these groups could add
$\tau_{int}$ = 0.4 - 1.9 cm$^{-1}$ to the 3.26 $\mu$m feature.  This
implies that 14 - 40 \% of the 3.26 $\mu$m band may be due to the PAH
CH stretch. This fraction could actually be much lower if a large
fraction of the PAHs would be frozen on the grains, causing the band
strength of the CH stretching mode to drop by a factor $\sim$ 4
(Joblin et al. 1994).  Indeed, it seems probable that in a high
density environment like the circumstellar regions of a young stellar
object a large fraction of highly refractory molecules such as PAHs
would accrete on the dust grains. In conclusion, it seems probable
that the contribution of PAHs to the 3.26 $\mu$m band is rather
small. This agrees with our result in Sect. 4.3, namely, that,
based on the assignment of the 6.85 $\mu$m band to $\rm NH_4^+$, most
of the 3.26 $\mu$m feature should be caused by the $\rm NH_4^+$ $\nu_2+\nu_4$
band.

Contrary to our result it was concluded by Bregman et al. (2000) that
PAHs may account for the entire 3.26 $\mu$m absorption band.  Some
important differences between his calculations and ours must be
pointed out. First, the integrated intensity of the 3.26 $\mu$m band
was approximated by depth x FWHM in this earlier paper, while we use
exact numerical integration, arriving at a 1.7 times higher
intensity. Second, Bregman et al. estimated that typically only 17 -
30 \% of the entire emission in the CH bending region originates from
the 11.2 $\mu$m band. However, a thorough recent survey of this region
in a wide variety of objects (Hony et al. 2001) showed that this
fraction is considerably higher, i.e., 30 - 70 \%.

\subsection{The abundance of $\rm NH_4^+$ and the counterions}

Applying the intrinsic strength of the $\nu_4$ $\rm NH_4^+$ band, the
$\rm NH_4^+$ column density can be derived from the observations.
With A($\rm NH_4^+$) = 4.4 $\times$ $\rm 10^{-17}$ cm ion$^{-1}$ 
(Sect. 3.3; uncertainty factor 2), we find for
W33A, with an integrated optical depth of $\tau_{int}$(6.85 $\mu$m) = 
57 $\rm cm^{-1}$ (after
correction for the $\rm CH_3OH$ contribution), 
N($\rm NH_4^+$) = 1.3 $\times$ $\rm 10^{18}$ $\rm cm^{-2}$.  For W33A
we adopt N($\rm H_2O$) = 1.2 $\times$ $\rm 10^{19}$
$\rm cm^{-2}$. This number was derived by Gibb et al. 2000
by fitting the blue wing of the 3 $\mu$m band with laboratory ice bands.
It must be noted that those authors adopted a
slightly lower column density as their end result, based, in part, 
on matches of the blue wing of the 3 $\mu$m band of W33A with
other interstellar 3 $\mu$m features. However, the correspondence
of the W33A 3 $\mu$m band with the other interstellar features was 
considerably less than the correspondence with the selected laboratory
ice bands (presumably caused by source to source variations in ice 
composition and temperature). Therefore, we prefer to adopt here the value
which is solely based on the laboratory comparison. This gives an
$\rm NH_4^+$ abundance of 11 \% for W33A, with an uncertainty of a factor
2 due to the bandstrength error. 

From the data summarized in Table~\ref{featuresobs}, it can be derived 
that the $\rm
NH_4^+$ abundance varies from 3 \% to 17 \% relative to $\rm
H_2O$. Here we use a constant correction for the contribution of
methanol to the 6.85 $\mu$m band of 20 \%, as for W33A.  This is
perhaps a bit large, since in general the methanol abundance in
YSO's is lower than for W33A (Dartois et al. 1999b).  However, the
correction is small anyway, and no attempt was made for a detailed
source to source analysis.  One exception is GL7009S, which has an
exceptionally high abundance of solid methanol, for which we adopted a
30 \% correction of the 6.85 $\mu$m feature (Dartois et al. 1999a).

Clearly, an equal amount of negative charge should be present in the
ices to balance $\rm NH_4^+$. Recently, Gibb et al. (2000)
investigated the full inventory of ice species
along the line-of-sight towards the high mass YSO W33A. Their analysis
showed that the negative species that have been detected towards this source,
i.e., $\rm OCN^-$ and $\rm HCOO^-$ (assuming that $\rm OCN^-$ is
the carrier of the interstellar XCN feature; Grim \& Greenberg 1987;
Schutte \& Greenberg 1997; Demyk et al. 1998; Hudson \& Moore 2000; 2001; 
Novozamsky et al. 2001)
have abundances of 3.2 \% and 0.8 \%, respectively (these values are
slightly lower than those of Gibb et al., due to the 
10 \% higher $\rm H_2O$ column we adopt for W33A). Since the
$\rm NH_4^+$ abundance for this source is $\sim$11 \%, this implies that only 
$\sim$ 30 \% of the countercharge is provided
by the observed negative species. Our experiments suggest that the
residual charge could be provided by a combination of 
ions such as $\rm HCO_3^-$, $\rm NO_2^-$, $\rm NO_3^-$. As was shown,
a mixture of these species embedded in an $\rm H_2O$-dominated ice
does not produce significant spectral structure. Their total
abundance would be $\sim$ 7 \% of $\rm H_2O$ for W33A. For other sources, the
fraction of the charge balance that can be provided
by the observed negative species (i.e., $\rm OCN^-$) is even smaller 
(Gibb et al. 2000). 
Thus, assuming that they make up for the residual balance, 
the total abundance of ``invisible'' anions in the ices near 
YSO's would be 
in general $\sim$ 70 - 100 \% of $\rm NH_4^+$, i.e., 3 - 15 \%.

It has been argued that the poor correlation between the 4.62 $\mu$m
feature of $\rm OCN^-$ and the 6.85 $\mu$m band disagrees with an
identification of this feature with $\rm NH_4^+$ (Keane et al. 2001;
cf. Figs.~\ref{W33A} \& ~\ref{MonR2comp3p26}. However, as pointed 
out above, 
$\rm OCN^-$ provides only a small fraction ($\leq$ 20 \%) of the
countercharge. In such a case, a clear correlation between the
$\rm OCN^-$ and $\rm NH_4^+$ abundances is only expected if the
make-up of the anion mixture is stable between YSO's. The lack
of correlation shows that the fraction of $\rm OCN^-$ in the total
ensemble of anions strongly varies.

\section{Astrophysical Implications}

When the $\rm NH_4^+$ assignment was originally made (Grim et
al. 1989a), the identity of the counterions was uncertain. This was
widely considered a prime concern (Schutte et al. 1996a; Keane et
al. 2001; Demyk et al.  1998; Tielens and Whittet 1997). Our new
experiments demonstrate that in $\rm H_2O$ dominated ices
the charge balance can be achieved without introducing spectral
structure of negative ions
that is inconsistent with the observations. Thus it appears that the
objections to the $\rm NH_4^+$ assignment have been met. All things
considered, $\rm NH_4^+$ now stands as the best assignment for the
interstellar 6.85 $\mu$m feature (apart form the generally small fraction
caused by $\rm CH_3OH$; Sect. 4), based on the band's position, shape,
and width. Moreover, the ammonium ion matches both the 6.85 and 3.26
$\mu$m features, and is able to reproduce the temperature-dependent
redshift of the 6.85 $\mu$m band. Chemical considerations also support
the assignment as $\rm NH_4^+$ is easily produced by either photolysis
or ion irradiation of known, or suspected, interstellar molecules, or
by mild warm-up of ammonia and acids which could be present in the ice
(see below).  A straightforward observational test would be to probe
the tight correlation between the 3.26 and the 6.85 $\mu$m features
which is expected if both are due to $\rm NH_4^+$.

It is unclear whether interstellar $\rm NH_4^+$ could arise from
irradiation or photolysis as in our experiments. The weakening of
the 3.4 $\mu$m CH stretching mode of aliphatic grain material
in goig from diffuse to dense regions indicates that energetic 
processing could play an important role in
dense regions (Mu\~noz Caro et al. 2001; Mennella et al. 2001).
However, the processing of our samples gives rise to some weak
features which have not yet been observed, most notably the $\rm O_3$
band at 1038 $\rm cm^{-1}$, and the unidentified 1270 $\rm cm^{-1}$
feature (Fig.~\ref{comp4part}). Neither of these bands are visible in the
ISO-SWS observations of YSO's (Fig.~\ref{W33Acomp}; Gibb et
al. 2001).  Alternatively, the formation of $\rm NH_4^+$ may not be
associated with energetic processing at all. Perhaps $\rm NH_3$ and
acids like $\rm H_2CO_3$, $\rm HNO_3$, $\rm HNO_2$, HNCO or even $\rm
H_2SO_3$ and $\rm H_2SO_4$ can be formed by surface chemistry. Whether
such relatively large species are formed by surface reactions depends
on the extent at which the exothermicity of the reactions can be
channeled into translational energy of the products (Brown 1990;
Hasegawa et al. 1992). Acid-base reaction and ion formation could also
be part of the surface reaction sequence, or be promoted by warm-up or
a small amount of radiation (Novozamsky et al. 2001, van Broekhuizen
et al. 2003).

While in this view the anions would be undetectable in the solid state,
possibly the corresponding acids could be found in the gas phase 
in the hot cores of star forming
regions where the ice mantles evaporate. A search for molecules
like $\rm H_2CO_3$, $\rm HNO_3$, $\rm HNO_2$ or $\rm H_2SO_3$ in such regions
seems therefore warranted.

From the limited sample of objects listed in Table~\ref{featuresobs}
it is unclear whether $\rm NH_4^+$ has different abundances for low
and high-mass objects. The low-mass protostar Elias 29 in the $\rho$
Oph cloud has an $\rm NH_4^+$ abundance of 3 \%, substantially less
than for the high mass sources.  However, the other low mass object in
our sample, HH100, gives a relatively high abundance of 8 \% (albeit
with a large errorbar). More observations of the 6.85 $\mu$m feature
of low mass protostars and background field stars are desirable to
clarify which dense environments favor the formation of $\rm NH_4^+$.

In hot core regions where
elevated grain temperatures cause sublimation of the icy mantles,
large gas phase abundances of ammonia have been found ((1 - 10)
$\times$ 10$^{-6}$ relative to hydrogen; Blake et al. 1987; Heaton et
al. 1989; Cesaroni et al. 1994). This considerably exceeds the
abundance of $\rm NH_3$ in cold dense regions ($\rm N(NH_3)$/$\rm
N(H_2)$ $\approx$ 10$^{-7}$; Federman et al. 1990), indicating that
ice sublimation is an important source of $\rm NH_3$. However, recent
observations failed to detect the 2.21 $\mu$m feature of $\rm NH_3$
towards W33A, indicating that the $\rm NH_3$ abundance in interstellar
ice may be low ($<$ 5 \%; Taban et al. 2003). A high abundance of
solid $\rm NH_4^+$ in interstellar ices could alleviate this dilemma,
since sublimation of this compound takes place during warm-up
upon a reverse acid-base reaction and the re-formation of $\rm NH_3$.

$\rm NH_4^+$ can only be formed from
$\rm NH_3$. The presence of solid $\rm NH_4^+$ thus indicates that 
solid $\rm NH_3$ was efficiently
formed during the ice accretion, presumably by hydrogenation of atomic
nitrogen on the grain surface (Hiraoka et al. 1995).  Whether in dense
regions gaseous nitrogen will reside in its molecular or its atomic
form depends on the depletion of oxygen, since reactions of atomic
nitrogen with OH is an essential step towards $\rm N_2$ formation
(Charnley \& Rodgers 2002).  Thus the quantity of $\rm NH_4^+$ that is
present in the ices gives important information on the gas phase
conditions at the epoch of condensation. Thus probing the strength
of the 6.85 $\mu$m band could become
an invaluable tool for studying the chemical conditions associated
with the different regions in dense clouds, i.e., quiescent, as well
as near embedded high mass and low mass YSO's.

\section{Conclusions}

Experiments involving UV photolysis of $\rm
H_2O$/$\rm CO_2$/$\rm NH_3$/$\rm O_2$ ice mixtures efficiently produce
ammonium ($\rm NH_4^+$) together with probably $\rm HCO_3^-$, $\rm
NO_3^-$, and $\rm NO_2^-$. These ions are produced by acid-base reactions
between the $\rm NH_3$ and photochemically produced acids. 
The $\nu_4$ mode of $\rm NH_4^+$ gives a strong
feature at 6.85 $\mu$m.  Other $\rm NH_4^+$ features fall at
3.27 and 3.50 $\mu$m. Due to strong interaction of
the ion with the matrix, the $\nu_4$ $\rm NH_4^+$ band shifts strongly
redward with increasing temperature. The infrared signature of the 
counterions is very weak for $\rm H_2O$-dominated ices. 

Comparison to the 6.85 $\mu$m feature towards embedded young stellar
objects shows that the $\nu_4$ $\rm NH_4^+$ feature obtained in our
experiments provides a good match.  The observed red shift of the
interstellar band with increasing line-of-sight dust temperature is
also consistent with this assignment.  Furthermore, features are
observed at 3.26 and 3.48 $\mu$m towards these objects which could be
(partly) due to the ammonium ion.  The implied abundance of $\rm
NH_4^+$ relative to $\rm H_2O$ is 3 - 17 \%.

The two
anions observed in interstellar ices, $\rm OCN^-$ and (possibly) $\rm HCOO^-$,
are insufficiently abundant to balance the positive charge from the
$\rm NH_4^+$ present.  Therefore heavier negative ions of the kind
formed in our experiments appear to be required. It is yet unclear
whether such ions are formed by energetic processing, as in the
laboratory experiments, or by grain surface chemistry.

Future laboratory studies should include analysis by mass 
spectroscopy to exactly
define the nature of the photochemically produced acids. Sulfur 
could be added to the mixture to extend the number of acids
that are produced. On the observational side, the $\rm NH_4^+$
assignment could be tested by probing the correlation between the 3.26
and 6.85 $\mu$m bands.  In addition, it would be very interesting to
search for rotational emission bands of evaporating acids like $\rm
HNO_2$, $\rm H_2CO_3$ or $\rm H_2SO_3$ in hot core regions.

\begin{acknowledgements}
First of all, we thank Marla Moore and Reggie Hudson for their
invaluable partnership in this project.  We thank Mayo Greenberg, Ruud
Grim, Guillermo Mu\~noz Caro, Richard Ruiterkamp, and Ewine van
Dishoeck for stimulating discussions. We furthermore thank Oswin
Kerkhof and Jacqueline Keane for their help with the experimental
effort. Finally, we thank Erika Gibb, Kris Sellgren, Adwin Boogert and
T.  Brooke for making the electronic version of their data available
to us.  The paper greatly benefitted from the comments of an anonymous
referee.  Careful proofreading of the manuscript by Fleur van
Broekhuizen was a great help. One of us (R.K.K.) acknowledges the
support of the NASA Goddard Space Flight Center.

\end{acknowledgements}

\end{document}